\documentclass[11pt,a4paper]{article}
\usepackage{jheppub}
\usepackage{amscd}
\usepackage{amsfonts}
\usepackage{amsthm}
\usepackage{stmaryrd}
\usepackage[normalem]{ulem}
\usepackage{tikz}
\usetikzlibrary{arrows,shapes}

\def\IZ{\mathbb{Z}}

\def\IX{\mathbb{X}}

\def\IB{\mathbb{B}}

\newcommand{\Tr}{\operatorname{Tr}}

\newcommand{\pp}{{=\!\!\!|}}
\newcommand\fverb{\setbox\pippobox=\hbox\bgroup\verb}
\newcommand\fverbdo{\egroup\medskip\noindent%
            \fbox{\unhbox\pippobox}\ }
\newcommand\fverbit{\egroup\item[\fbox{\unhbox\pippobox}]}
\newbox\pippobox

\title{Aspects of the Doubled Worldsheet}

\author[a]{Sibylle Driezen,}
\author[a,b]{Alexander Sevrin,}
\author[a]{ and Daniel C. Thompson}
\affiliation[a]{
Theoretische Natuurkunde, Vrije Universiteit Brussel\\  \& The International Solvay Institutes\\
Pleinlaan 2, B-1050 Brussels, Belgium }
\affiliation[b]{also at the Physics Department, Universiteit Antwerpen\\
Campus Groenenborger, 2020 Antwerpen, Belgium}

\emailAdd{Sibylle.Driezen@vub.ac.be}
\emailAdd{Alexandre.Sevrin@vub.ac.be}
\emailAdd{Daniel.Thompson@vub.ac.be}

\abstract{We clarify the relation between various approaches to the manifestly T-duality symmetric string. We explain in detail how the PST covariant doubled string arises from an unusual  gauge fixing. We pay careful attention to the role of  ``spectator'' fields in this process and also show how the T-duality invariant doubled dilaton emerges naturally.   We   extend these ideas to non-Abelian T-duality and show they give rise to the duality invariant formalism based on the semi-Abelian Drinfeld Double.  We then develop  the ${\cal N}=(0,1)$ supersymmetric duality invariant formalism.  }


\begin{document}
 
\maketitle

\setcounter{equation}{0}

\section{Introduction}
 
A central theme in recent years has been to understand the ways in which dualities of string and M-theory may be promoted to manifest symmetries and indeed the extent to which they may be used to determine the structure of the underlying theory.

The idea of a T-duality invariant worldsheet description, {\em the doubled worldsheet} ({\bf DWS}), of strings goes back to pioneering work of Duff  \cite{Duff:1989tf}  and Tseytlin \cite{Tseytlin:1990nb,Tseytlin:1990va}.   The study of this approach was reignited following the proposal of Hull \cite{Hull:2004in,Hull:2006va} to use such a formalism to define strings in a class of non-geometric backgrounds known as T-folds.  Parallel to this has been the development of a spacetime T-duality invariant theory, often now dubbed {\em double field theory} ({\bf DFT}),  whose origins date to the seminal works of Tseytlin  \cite{Tseytlin:1990nb,Tseytlin:1990va} and Siegel  \cite{Siegel:1993xq,Siegel:1993th}.  This approach was derived from  the perspective of closed string field theory on a torus by Hull and Zwiebach \cite{Hull:2009mi}.  These ideas have also been explored in the context of M-theory \cite{Berman:2010is,Hohm:2013pua} where {\em exceptional field theory}  ({\bf EFT})  seeks to promote the U-duality group to a manifest symmetry of a spacetime action and in the $E_{11}$ program of West \cite{West:2001as} and collaborators.\footnote{Our focus in this note will be on the worldsheet rather than spacetime so for further introduction to the  {\bf DFT} and  {\bf EFT}  we refer the reader to the review articles \cite{Aldazabal:2013sca,Hohm:2013bwa,Berman:2013eva}.}  

A common theme of the {\em doubled worldsheet}, {\em double field theory} and {\em exceptional field theory} is that in order to make the duality act as a linearly realised symmetry, the dimensionality of spacetime is augmented by the introduction of additional coordinates.  For instance, in the case of T-duality of strings on a $d$-dimensional torus, we have a $2d$-dimensional extended spacetime consisting of $d$-regular coordinates $x^i$ and $d$-dual coordinates $\tilde{x}_i$.  Just as position is conjugate to momenta one can think of these extra coordinates as conjugate to winding of the string.  For strings in a curved background, the components of the background  metric $g_{ij}$ and NS two-form fields $b_{ij}$ in the internal toroidal directions  are united into a {\em generalised metric}, 
\begin{equation}
{\cal H}_{IJ}=
\left(
\begin{array}{cc}
g-b\,g^{-1}\,b&-b\,g^{-1}\\
g^{-1}\,b&g^{-1}\end{array}
\right)\,,
\end{equation}
on the doubled space parametrised by coordinates $\mathbb{X}^I = \{ x^i , \tilde{x}_i \}$.   The T-duality group, which is $O(d,d;\mathbb{Z})$ in this case, acts on this {\em generalised metric} as, 
\begin{equation}
{\cal H}\rightarrow {\cal H}'= {\cal O}^T {\cal H} {\cal O }\, ,
\end{equation}
where the group element preserves the inner product, ${\cal O}^T \,\eta\, {\cal O} = \eta $,  given in this basis by, 
\begin{equation}
\eta_{IJ} =  \left(
\begin{array}{cc}
0&{\bf 1}\\
{\bf 1}&0\end{array}
\right)\,.
\end{equation}
From the generalised metric we see that the doubled space is equipped with an almost product structure ${\cal S}= \eta {\cal H}$ such that ${\cal S}^2 = {\bf 1}$ which one can think of as giving rise to a ``chiral structure'' specified by the projection operators,
\begin{eqnarray}
P_\pm\,=\,\frac 1 2 \,\big({\bf 1}\pm S\big)\,.
\end{eqnarray}
This doubled space is also equipped with a natural symplectic product $\Omega$ given in this basis as,
\begin{equation}
\Omega_{IJ} =  \left(
\begin{array}{cc}
0&{\bf 1}\\
-{\bf 1}&0\end{array}
\right)\,.
\end{equation}
The existence of the objects ${\cal S}$, $\eta$ and $\Omega$ are central to the recent proposal of Born geometries \cite{Freidel:2013zga,Freidel:2015pka}.

In all of these duality symmetric approaches, there is a  price to pay; action principles based on the doubled or extended spacetimes require  supplementary constraints.  In {\bf DFT} and {\bf EFT} gauge invariance of the theory requires a constraint, also known as the section condition, that essentially declares the field content of the theory to depend on only a physical spacetime's worth of coordinates.\footnote{Upon solving this section condition for type II DFT  \cite{Hohm:2011dv} globally, one recovers the generalised geometry \cite{Gualtieri:2003dx,Hitchin:2004ut} reformulation of supergravity of \cite{Coimbra:2011nw,Coimbra:2012af}. Ways in which the section condition can be consistently relaxed are of great interest and connect to gauged supergravities see \cite{Hohm:2011cp,Geissbuhler:2011mx,Aldazabal:2011nj,Grana:2012rr,Berman:2012uy}. }     In the case of {\bf DWS}, the worldsheet bosons are required to obey a chirality constraint meaning that of the $2d$ bosons $\mathbb{X}^I$ exactly half are left-movers and half are right-movers and thereby give the correct contribution to the physical central charge.  In this note we will be focussed on the variety of ways in which the chirality constraint of {\bf DWS} has been handled.  

To explain this let us momentarily restrict ourselves to a simple case; a doubled torus $T^{2d}$ with coordinates $\mathbb{X}^I$ and a generalised metric $\cal H$ possibly depending on some other ``spectator'' coordinates $y$ parametrising a base manifold over which the doubled torus is trivially fibered with vanishing connection.  In this situation the constraints are given in terms of the chiral projections by,
\begin{equation}
\label{eq:Chirality}
(P_+)^I{}_J \partial_= \mathbb{X}^J = 0  \ , \qquad  (P_-)^I{}_J \partial_\pp\, \mathbb{X}^J = 0  \ . 
\end{equation} 
Chiral scalars are notoriously tricky objects to describe, the main reason is that these constraints are first order differential equations and in the terminology of Dirac second class constraints and can not be imposed easily with Lagrange multipliers.  One approach is to simply consider a non-linear $\sigma$-model in the doubled spacetime,
\begin{equation}
\label{eq:Hull}
S_{\mbox{\footnotesize Hull}} = \frac 1 2  \int d^2 \, \sigma \,\partial_\pp \, \mathbb{X}^I\,{\cal H}_{IJ}\,  \partial_=  \mathbb{X}^J + \dots  \ , 
\end{equation} 
in which, and in the following, the ellipsis indicate terms depending on the spectators and also a topological term involving $\Omega_{IJ}$ both of which we shall detail later.  One can then implement the constraints supplementary to the action for instance by using Dirac brackets and then performing canonical quantisation \cite{HackettJones:2006bp} or by holomorphic factorisation of the resulting partition function \cite{Berman:2007vi,Tan:2014mba}.   Whilst this is certainly a viable route, one should very much like to have an action principle from which eq.~\eqref{eq:Chirality} follows.  Without introducing   extra field content this is possible only at the expense of sacrificing manifest Lorentz invariance leading to the action pioneered by Tseytlin  \cite{Tseytlin:1990nb,Tseytlin:1990va},
\begin{equation}
\label{eq:Tseytlin}
S_{\mbox{\footnotesize Tseytlin}} =  \frac 1 4  \int d^2 \sigma\,  - \partial_\sigma  \mathbb{X}^I \,{\cal H}_{IJ} \,\partial_\sigma  \mathbb{X}^J + \partial_\sigma \mathbb{X}^I\,\eta_{IJ} \, \partial_\tau \mathbb{X}^J+  \dots \ , 
\end{equation} 
which essentially employs a Floreanini-Jackiw \cite{Floreanini:1987as} construction for chiral bosons.  The equations of motion that follow from eq.~\eqref{eq:Tseytlin} may be integrated and using a gauge invariance of the form $\delta X^I = f^I(\tau)$ give rise to the desired chirality constraints of  eq.~\eqref{eq:Chirality}.   Despite its apparent non-covariance one can still employ some conventional field theory techniques, for instance one-loop beta functions of this action have been calculated \cite{Berman:2007xn,Berman:2007yf} and shown to give rise to background field equations for ${\cal H}$ which are indeed compatible with the equations that follow from {\bf DFT} in the present context (other attempts to make more precise the linkage between {\bf DFT} and the worldsheet theory by allowing ${\cal H}$ to depend on the internal coordinates are found in \cite{Copland:2011wx}, \cite{Lee:2013hma} and  \cite{Betz:2014aia}).    However multi-loop calculations are at best very difficult without Lorentz covariance.  

A further approach is to include extra fields so as to furnish the action with a gauge redundancy which promotes the second class constraint to a first class one.  This is the spirit of the Pasti-Sorokin-Tonin (PST) approach to chiral fields\footnote{A different approach based upon gauging the the symmetries generated by the constraints was followed in \cite{Hull:2006va} however at the cost of loosing manifest $O(d,d;\IZ)$ invariance.  This approach can be extended to superspace (at least to ${\cal N}=(1,1)$) and higher genus worldsheets.  } \cite{Pasti:1996vs}.  In the present context this leads to a doubled action,  
\begin{equation}
\label{eq:PST}
\begin{aligned}
S_{\mbox{\footnotesize PST}} =&  \frac 1 2  \int d^2 \sigma\  \partial_\pp\,  \mathbb{X}^I\,{\cal H}_{IJ}\, \partial_=  \mathbb{X}^J  -  \frac{\partial_\pp\, f}{\partial_= f }\, (P_+ \partial_=  \mathbb{X})^I  \eta_{IJ}(P_+ \partial_=  \mathbb{X})^J   \\ 
&\qquad \qquad 
+ \frac{\partial_= f}{\partial_\pp\, f}\,  (P_- \partial_\pp\,  \mathbb{X})^I \eta_{IJ}(P_- \partial_\pp \, \mathbb{X})^J   + \dots \ . 
\end{aligned}
\end{equation} 
The symmetries of this action are, 
\begin{eqnarray}
\label{eq:symmetries}
1)& & \qquad  \delta \mathbb{X}^I = \Lambda^I(f)   \ , \quad \delta f = 0 \ ,\\ 
2) &&  \qquad \delta \mathbb{X}^I =  \frac{\epsilon}{\partial_\pp\, f}  (P_- \partial_\pp\,  \mathbb{X})^I  + \frac{\epsilon}{\partial_= f}  (P_+ \partial_=  \mathbb{X})^I   \ , \quad \delta f = \epsilon \ . 
\end{eqnarray}
Upon using the second symmetry to fix $f= f(\tau)$ one recovers the Tseytlin action  eq.~\eqref{eq:Tseytlin}. 

Whilst this overall picture is correct, the literature has been rather sketchy in places about many of the details concerning the derivations of the covariant forms of the doubled worldsheet and in particular has omitted a careful treatment of spectator fields (i.e. exactly the terms in ellipsis in the above discussion).   In the following we will resolve many of these outstanding issues and by giving the complete derivation of the covariant bosonic doubled formalism achieved by adopting an unusual gauge fixing in a Buscher procedure.  A version of this idea was suggested in \cite{Rocek:1997hi} wherein an axial gauge fixing  gives rise to the non-covariant action eq.~\eqref{eq:Tseytlin} and more recently explored in \cite{Nibbelink:2012jb,Nibbelink:2013zda} wherein covariant gauge fixing choices were adopted (though those are not directly relevant to the present discussion).  We further this approach by making direct the linkage to the PST form of the action and will then be able to clarify some surprising features concerning the origin of the PST symmetries.   These ideas will then be generalised to the case of non-Abelian T-dualities \cite{de la Ossa:1992vc} and we will recover a covariant version of the Poisson-Lie duality symmetric action of Klim\v{c}\'ik and \v{S}evera \cite{Klimcik:1995ux}, \cite{Klimcik:1995dy}.  

This work arose out of an ongoing attempt to better understand the supersymmetric doubled formalism. It is quite clear in this case how to generalise the chirality constraints to ${\cal N}=1$ supersymmetry; one promotes partial derivatives to super covariant derivatives acting on superfields, 
\begin{equation}
\label{eq:sChirality}
(P_+)^I{}_J D_- \mathbb{X}^J = 0  \ , \qquad  (P_-)^I{}_J D_+ \mathbb{X}^J = 0  \ . 
\end{equation} 
Previous work in the literature has followed the route of imposing the constraints by hand either via Dirac brackets as in \cite{HackettJones:2006bp} or via holomorphic factorisation of a partition function \cite{Chowdhury:2007ba}.   However  the implementation of these constraints at the level of the action has rarely been considered; there is no known covariant formalism in the style of eq.~\eqref{eq:PST} and even a non-covariant Tseytlin style action has only been considered for the case of a constant generalised metric i.e. assuming no dependance on spectator coordinates.   Naive attempts to generalise this appear to fail badly and addressing this short coming seems essential for the doubled worldsheet to have a life in superstring theory. 

Here we take a step in this direction by carefully analysing the simplest supersymmetric model, {\em i.e.} the one which exhibits an ${\cal N}=(0,1)$ supersymmetry on the worldsheet. It turns out that even a first order formulation comes accompanied by external constraints. These extra constraints are similar in nature to nilpotency constraints on superfields, so at the level of the components of the superfields they are algebraic and, as a consequence, they can simply be imposed using Lagrange multipliers. We do give both the PST and the Tseytlin like description of the ${\cal N}=(0,1)$ system. Note that  a Hamiltonian perspective was given in  \cite{Blair:2013noa} in which only bosonic degrees of freedom are doubled making supersymmetry less evident; here we will instead work in superspace. 

\section{Bosonic Abelian Doubled String} 
\subsection{Deriving the covariant doubled string }
Our starting point is some compact $D$-dimensional manifold ${\cal M}$ endowed with a metric $g$ and a closed 3-form $H$. Locally we introduce the Kalb-Ramond 2-form $b$: $H=db$. Choosing local coordinates $X^A$, $A\in\{1,\cdots,D\}$, the non-linear $\sigma $-model Lagrange density is given by,
\begin{eqnarray}
{\cal L}= \partial _\pp\, X^A\big(g_{AB}+b_{AB}\big) \partial _= X^B\,.
\end{eqnarray} 
We now assume the existence of $d$ isometries ($d\leq D$) and introduce adapted coordinates $x^i$, $i\in\{1,\cdots,d\}$, such that the background fields $g$ and $b$ {\em do not depend on} $x$. The spectator coordinates are called $y^\alpha $, $\alpha \in\{1,\cdots,D-d\}$. In an obvious matrix notation the Lagrange density becomes,
\begin{eqnarray}
{\cal L}= \partial _\pp\, x^T\,E \,\partial _=x+ \partial _\pp\, x^T\,M \,\partial _=y+\partial _\pp\, y^T\,N \,\partial _=x+
\partial _\pp\, y^T\,K \,\partial _=y\,.\label{l1}
\end{eqnarray}
A special role is accorded to $E_{ij}=g_{ij}+b_{ij}$. We denote the inverse of $g_{ij}$ ($g$) by $g^{ij}$ ($g^{-1}$): $g_{ik}\,g^{kj}= \delta _i^j$ ($g\,g^{-1}=g^{-1}\,g={\bf 1}$).
We introduce  ``connections'' ${\cal B}$ and $\tilde {\cal B}$,
\begin{eqnarray}
{\cal B}^i&=&g^{ij}g_{j \beta }\, \partial y^\beta \, , \nonumber\\
\tilde{\cal B}_i&=&b_{i \beta } \,\partial y^\beta -b_{ij}\,g^{jk}g_{k \beta } \partial\, y^\beta\, ,
\end{eqnarray}
 which are adapted coordinate representations of (pull backs of) one-forms detailed in  \cite{Hull:2006va}  that are  horizontal and invariant  with respect to the Killing vectors generating the isometry.  With these we may rewrite eq.~(\ref{l1}) as,
\begin{eqnarray}
{\cal L}= \nabla_\pp\,x^T\,E\,\nabla_=\,x+ \partial _\pp\,x^T\,\tilde {\cal B}_=-\partial _=\,x^T\,\tilde {\cal B}_\pp - {\cal B}_\pp^T\,E\, {\cal B}_=+
\partial _\pp\, y^T\,K \,\partial _=y\,,
\end{eqnarray} 
where,
\begin{eqnarray}
\nabla x= \partial  x+ {\cal B}\,.
\end{eqnarray}
In order to obtain the T-dual model we gauge the isometries,
\begin{eqnarray}
x \rightarrow x'=x+ \zeta\,,\qquad y \rightarrow  y'=y\,.\label{gs1}
\end{eqnarray}
For this we introduce $2d$ gauge fields $A_\pp$ and $A_=$ transforming as,
\begin{eqnarray}
A_\pp \rightarrow A_\pp'=A_\pp- \partial _\pp\, \zeta \,,\qquad
A_= \rightarrow A_='=A_=- \partial _=\, \zeta \,,\label{gs2}
\end{eqnarray}
together with $d$ Lagrange multipliers $\tilde x$ which are inert under the gauge transformations. The gauge invariant Lagrange density is then\footnote{Note that in order to avoid nontrivial holonomies around non-contractible loops, $\tilde x$ should satisfy appropriate periodicity conditions. In addition,    a surface term $ \partial _=(\tilde x^TA_\pp)-\partial _\pp(\tilde x^TA_=)$ should be added to eq.~(\ref{l2}) \cite{Rocek:1991ps} which is important to keep in mind as we treat boundary contributions in what follows. },
\begin{eqnarray}
{\cal L}_{\mbox{\footnotesize gauged}}&=& \big(\nabla_\pp\,x+A_\pp\big)^T\,E\,\big(\nabla_=\,x+A_=\big)+ \big(\partial _\pp\,x+A_\pp\big)^T\tilde {\cal B}_=-\big(\partial _=\,x+A_=\big)^T\tilde {\cal B}_\pp \nonumber\\
&&- {\cal B}_\pp^T\,E {\cal B}_=+
\partial _\pp\, y^T\,K \,\partial _=y+\tilde x^T\,\big(\partial _\pp\,A_=-\partial _=A_\pp\big)\,.\label{l2}
\end{eqnarray}
Integrating over the Lagrange multipliers sets the field strengths to zero, so we can gauge away the gauge fields and we recover the original model.    
However making the gauge choice $x=0$, integrating by parts on the Lagrange multiplier term  and integrating out the gauge fields yields the dual model,
\begin{eqnarray}
\tilde {\cal L}_{\mbox{\footnotesize dual}} = \nabla_\pp\,\tilde x^T\,\tilde E\, \nabla_=\,\tilde x+ \partial _\pp\,\tilde x^T\, {\cal B}_=-\partial _=\,\tilde x^T\, {\cal B}_\pp -\tilde  {\cal B}_\pp^T\,\tilde E\, \tilde {\cal B}_=+
\partial _\pp\, y^T\,\tilde K \,\partial _=y\,,
\end{eqnarray} 
with,
\begin{eqnarray}
 \nabla \tilde x= \partial  \tilde x+ \tilde {\cal B}\,.
\end{eqnarray} 
The dual background fields are given by the Buscher rules \cite{Buscher:1987sk}, \cite{Buscher:1987qj},
\begin{eqnarray}
\tilde E = E^{-1}\,,\qquad \tilde M= E^{-1}M\,,\qquad \tilde N=-NE^{-1}\,,\qquad \tilde K= K-NE^{-1}M\,,
\end{eqnarray}
together with a shift in the dilaton that is seen when the dualisation procedure is carried out in a path integral. 

Let us now turn to the manifest T-dual invariant or doubled formulation of the model. In fact our starting point, the gauged Lagrange density eq.~(\ref{l2}) is already ``doubled'' as both the original coordinates $x$ and the dual coordinates $\tilde x$ appear. This was suggested in \cite{Rocek:1997hi}
 (see also  \cite{Patalong:2013iya} for a detailed development) where it was shown that by making the non-Lorentz covariant gauge choice $A_\pp\,=A_=\equiv A$ and subsequently integrating out $A$ 
one recovers Tseytlin's non-Lorentz covariant doubled formulation \cite{Tseytlin:1990nb}, \cite{Tseytlin:1990va}. This is very reminiscent of the Floreanini-Jackiw formulation of a chiral boson \cite{Floreanini:1987as}. Just as the Floreanini-Jackiw formalism can be covariantized \cite{Pasti:1996vs}, \cite{Sevrin:2013nca} we expect the same for Tseytlin's action. In the next we show how by making a judicious gauge choice in eq.~(\ref{l2})  one indeed obtains a Lorentz invariant doubled worldsheet formulation.

Starting from the gauge invariant Lagrange density in eq.~(\ref{l2}) we impose the gauge fixing condition,
\begin{eqnarray}\label{eq:fixing}
\partial _\pp\,fA_= = \partial _=fA_\pp\,,
\end{eqnarray} 
where $f$ is some scalar field. In writing the gauge fixing choice as in eq.~\eqref{eq:fixing} we are emphasising that the function $f$ should be suitably chosen so as to have nowhere vanishing derivatives -- we will discuss this requirement further in the discussion section. 
 Making a coordinate transformation,
\begin{eqnarray}
\sigma ^\pp \rightarrow  \hat \sigma ^\pp=\sigma ^\pp\,,\qquad 
\sigma ^= \rightarrow  \hat \sigma ^==f\,,
\end{eqnarray}
the above gauge choice simplifies to $\hat A_\pp\,=0$. 
From this we immediately identify the residual gauge symmetry. It is given by eqs.~(\ref{gs1}) and (\ref{gs2}) where the gauge parameter $\xi$ is of the form $\xi=\xi(f( \sigma ^\pp, \sigma^=))$. A full and detailed discussion of the residual symmetries will be given in section 2.2. In addition to this, one verifies  that the Lagrange density eq.~(\ref{l2}) is invariant under,
\begin{eqnarray}
\tilde x \rightarrow  \tilde x'=\tilde x +\tilde \xi (f)\,,\label{egs}
\end{eqnarray} 
as well.

Now the strategy is clear. We adopt this gauge choice and parameterize the gauge fields as,
\begin{eqnarray}
A_\pp= A\, \partial _\pp\,f\,\qquad A_=\,= A\, \partial_=f\,,
\end{eqnarray}
where $A$ is a $d \times 1$ column matrix of scalar fields.
 Implementing this gauge fixing in eq.~(\ref{l2})  and eliminating $A$ through its equations of motion,
 \begin{eqnarray}
A=- \frac{1}{2 \partial _=f}\,g^{-1}\, {\cal J}_=
- \frac{1}{2 \partial _\pp\,f}\,g^{-1}\,{\cal J}_\pp \,, \label{E2O} 
\end{eqnarray}
where,
\begin{eqnarray}
{\cal J}_= = E\, \nabla _=\,x+ \nabla _=\,\tilde x  \ ,  \quad 
{\cal J}_\pp = E^T \,\nabla _\pp\, x- \nabla _\pp\,\tilde x \, ,\label{AE2O} 
\end{eqnarray}
yields, after a little manipulation, the desired covariant doubled Lagrange density\footnote{Note that a Lagrange density somewhat similar to this one has been obtained in the context  of heterotic strings compactified on a Narain torus 
\cite{Cherkis:1997bx}. },
\begin{eqnarray}
\label{eq:Ldoubled}
{\cal L}_{\mbox{\footnotesize doubled}}&=&   \frac 1 2 \,\nabla_\pp\, \IX^T{\cal H}\nabla_=\,\IX -\frac{1}{2} \partial_\pp\, \IX^T \Omega \partial_=\, \IX
- \frac 1 2 \,   \frac{\partial _= f}{\partial _\pp \,f} \,\nabla_\pp\,\IX^T {\cal H}P_-\nabla_\pp\,\IX
\nonumber\\
&&- \frac 1 2 \,  \frac{\partial _\pp\,f}{\partial _= f} \,\nabla_=\,\IX^T {\cal H}P_+\nabla_=\,\IX  
+\frac 1 2 \, \partial _\pp\,\IX^T\eta\, \IB_=-\frac 1 2 \, \partial _=\,\IX^T\eta \,\IB_\pp+ \partial  _\pp\, y^T\hat K \partial _=\,y\,,
\nonumber\\
\label{l3}
\end{eqnarray}
where,
\begin{eqnarray}
\IB=
\left(
\begin{array}{c}
{\cal B}\\
\tilde {\cal B}\end{array} 
\right)\, , \quad    \nabla \IX = \partial  \IX+\IB\,.
\end{eqnarray}
This action is now (almost) manifestly invariant under global $O(d,d; \mathbb{R})$ transformations acting as,
\begin{equation}
{\cal H}\rightarrow {\cal H}'= {\cal O}^T {\cal H} {\cal O }  \ , \quad \IX \rightarrow {\cal O }^{-1} \IX \ , \quad \IB \rightarrow {\cal O }^{-1} \IB \ .
\end{equation}
This invariance is further reduced to $O(d,d;\mathbb{Z})$ by demanding that the periodicities of the coordinates $\IX$ are preserved \cite{Hull:2004in,Hull:2009mi}.  

Note that other than the initial integration by parts on the Lagrange multiplier term (and, see footnote 4, in a careful gauging any boundary terms from this are canceled by a boundary contribution),  we have not discarded any total derivatives in this manipulation and the topological term,  $\partial_\pp\, \IX^T \Omega \partial_=\, \IX$,  appears automatically.   One might be tempted to ignore such a piece however this term is vital for instance in getting a correct factorisation of the partition function   and in \cite{Hull:2006va} this topological term ensures invariance under certain large gauge transformations that are used to define the quantum theory (as originally emphasized in \cite{Giveon:1991jj}).  We will see later that when generalised to non-Abelian T-duality   it will no longer remain topological and rather play the role of a potential for a WZ term. Strictly speaking this topological term spoils the invariance of the action under $O(d,d;\mathbb{Z})$  unless ${\cal O}^T   \Omega  {\cal O }=  \Omega$.   Evidently the $GL(d, \mathbb{Z})$ subgroup of the duality group preserves $\Omega$,  but for the remaining components of $O(d,d;\mathbb{Z})$, namely B-field shifts and Buscher dualities, one needs to exercise more care.   Properly normalised this topological term    \cite{Hull:2006va} evaluates to the sum of products of winding numbers   around canonically dual cycles and  in a fixed winding sector evaluates to $\pi \mathbb{Z}$  contributing a sign in the path integral. 
B-field shifts have the effect of adding $2\pi {\mathbb Z}$ to this contribution and thus leave the path integral invariant.    For T-dualities that simply swap $n$ coordinates the coefficient of the topological term is multiplied by  $(-1)^{n}$ and again the path integral is invariant.

The Lagrange density governing the spectator coordinates is altered as well -- a fact often ignored in the literature. Indeed the $O(d,d;\IZ)$ non-invariant background field $K$ is replaced by $\hat K$ which {\em is} invariant and explicitly given by,
\begin{eqnarray}\label{eq:baseK}
\hat K=K-\frac 1 2 \, N\,g^{-1}\,M-\frac 1 4\, M^T\,g^{-1}\,M-\frac 1 4 \, N\,g^{-1}\,N^T\,.
\end{eqnarray}

  The action of parity is slightly non-standard. Since parity acts as ${\cal P}: \{\sigma^\pp, \sigma^=\} \rightarrow \{\sigma^=, \sigma^\pp \}  $ leaving the one-form gauge connection $A_\mu dx^\mu$ invariant,  we require that ${\cal P}: \{  x, \tilde{x} \}  \rightarrow \{ x, - \tilde{x}\}$ for the gauged Lagrangian to have definite parity.   In terms of the doubled space we have ${\cal P} : \IX^I   \rightarrow {\cal P}^I{}_J \IX^J$ with ${\cal P}^I{}_J = -( \Omega\eta)^I{}_J $.   In addition, for the term $E_{ij} \partial_\pp\, x^i \partial_= x^j$ to have definite parity we should also insist that ${\cal P}: b_{ij} \rightarrow - b_{ij}$ which implies that the generalised metric must transform as ${\cal P}: {\cal H} \rightarrow {\cal P}\cdot {\cal H} \cdot {\cal P}$.    Making use of the identity ${\cal P}\cdot {\cal \eta} \cdot {\cal P}= -\eta$  we see  ${\cal P}: (P_+ \partial_= \IX)^I \rightarrow ({\cal P}\cdot P_- \partial_\pp\, X)^I$  and thus both the Tseytlin and PST actions have  definite parity.

\subsection{Gauge symmetries and the origin of PST symmetry}
In this section we investigate the symmetries of the manifest $O(d,d;\IZ)$ invariant Lagrange density. 
Upon gauge fixing  the original gauge symmetry eqs. (\ref{gs1}) and (\ref{gs2}) and passing to the second order formalism, the residual gauge symmetry extended by the symmetry eq.~(\ref{egs}) is given by,
\begin{eqnarray}
\label{eq:residual}
\IX \rightarrow \IX'=\IX + \Lambda (f)\,,\qquad f \rightarrow f'=f\,,\label{ressym}
\end{eqnarray}
where $\Lambda (f)$ is a $2d\times 1$ column matrix of arbitrary functions of $f$. This explains the origin of the first of the symmetries of eq.~\eqref{eq:symmetries}. However the appearance of the PST symmetry which acts as,
\begin{eqnarray}
\delta f&=& \varepsilon \,, \nonumber\\
\delta \IX &=& \frac{ \varepsilon }{\partial _\pp\,f}\,P_-\, \nabla _\pp\,\IX+
\frac{ \varepsilon }{\partial _=\,f}\,P_+\, \nabla _=\,\IX\,,\label{PSTtrans}
\end{eqnarray}
in  the second order formulation eq.~(\ref{l3}) is quite mysterious. It looks as if it is unrelated to the original gauge symmetry eqs. (\ref{gs1}) and (\ref{gs2}). In the remainder we explain how the PST symmetry originates from  the gauged $\sigma $-model  in eq.~(\ref{l2}).

Given an infinitesimal vector, $\xi^\mu$, $\mu\in\{=\!\!\!\!\!|\,\,,=\}$, we introduce the variations,
\begin{eqnarray}
\delta A_\mu&=& {\cal L}_\xi\,A_\mu= \partial _\mu \big(\xi^\nu A_\nu\big)+\xi^\nu F_{\nu\mu}\,, \nonumber\\
\delta  x&=& -\xi^\mu A_\mu\,, \nonumber\\
\delta \tilde x&=&-\xi^\pp \left( {\cal J}_\pp + E^T A_\pp \, \right) + \xi^= \left( {\cal J}_= + E A_=  \right) \, ,
\end{eqnarray}
where $F_{\mu\nu}= \partial _\mu A_\nu- \partial _\nu A_\mu$, $ {\cal L}_\xi$ is the Lie derivative along $\xi$ and ${\cal J}$ was given in eq.~(\ref{AE2O}). One easily verifies that the gauged $ \sigma $-model in eq.~(\ref{l2}) is invariant under these transformations. This is not so surprising as these transformations can be rewritten as,
\begin{eqnarray}
\delta x&=& - \xi^ \mu A_ \mu \,, \nonumber\\
\delta A_\pp\,&=& \partial _\pp\,( \xi^ \mu  A_\mu )-\xi^=\, \frac{ \delta {\cal S} }{ \delta \tilde x}\,, \nonumber\\
\delta A_=\,&=& \partial _=( \xi^ \mu  A_\mu )+\xi^\pp\, \frac{ \delta {\cal S} }{ \delta \tilde x}\,, \nonumber\\
\delta \tilde x &=& -\xi^\pp\, \frac{ \delta {\cal S} }{ \delta A_=}+\xi^=\, \frac{ \delta {\cal S} }{ \delta A_\pp}\,,\label{eomsym}
\end{eqnarray}
where we introduced the action $ {\cal S}=\int d^2 \sigma {\cal L}_{\mbox{\footnotesize gauged}}$. So one sees that this is not a new symmetry: it is a combination of a (field dependent) gauge transformation eqs. (\ref{gs1}), (\ref{gs2}) with parameter $\zeta=-\xi^\mu A_\mu$ and a (trivial) equations of motion symmetry. 

However the situation changes when making the gauge choice $A_\pp\,= \partial _\pp\,f\,A$, $A_== \partial _=f\,A$. The residual gauge symmetry  is now,
\begin{eqnarray}
x \rightarrow x'=x + \varepsilon  (f)\,,\qquad f \rightarrow f'=f\,,\qquad A \rightarrow A'=A- \frac{d \varepsilon(f )}{df}\,,\label{ressymnn}
\end{eqnarray}	
and the symmetries in eq.~(\ref{eomsym}) survive provided we assign the following transformation rules to $A$ and $f$,
\begin{eqnarray}
\delta f&=& \xi^\mu \partial _ \mu f\,,\nonumber\\
\delta A&=& \xi^\mu \partial _ \mu A\,.
\end{eqnarray}
Introducing the parameters $ \varepsilon$ and $ \varkappa $,
\begin{eqnarray}
\varepsilon \equiv\xi^\pp\, \partial _\pp\,f+\xi^= \partial _=\,f\,, \qquad
 \varkappa   \equiv\xi^\pp\, \partial _\pp\,f-\xi^= \partial _=\,f\,,
\end{eqnarray}
one rewrites the transformation rules as,
\begin{eqnarray}
\delta f&=& \varepsilon \,, \nonumber\\
\delta x&=&- \varepsilon \,A\,, \nonumber\\
\delta \tilde x &=& \frac \varepsilon 2\,\Big(
- \frac{{\cal J}_\pp}{ \partial _\pp\,f}
+ \frac{{\cal J}_=}{ \partial _=\,f}+2b\,A
\Big)- 
\frac{\varkappa  }{2 \partial _\pp\,f \partial _=f}\, \frac{ \delta {\cal S}  }{ \delta  A}\,, \nonumber\\
\delta  A &=& \frac \varepsilon  2 \,\Big(
\frac{ \partial _\pp\, A}{ \partial _\pp\,f}+ \frac{ \partial _= A}{ \partial _= f}
\Big)+\frac{\varkappa  }{2 \partial _\pp\,f \partial _=f}\, \frac{ \delta {\cal S}  }{ \delta  \tilde x}\,.
\end{eqnarray}
So one sees that the symmetry parameterized by $ \varepsilon   $ corresponds to a genuine gauge symmetry while the one parameterized by $ \varkappa  $ is a trivial equations of motion symmetry. Eliminating $A$ through its equations of motion, eq.~(\ref{E2O}) the $\varkappa$ dependent term in $ \delta \tilde x$ drops out and the transformations rules for $x$, $\tilde x$ and $f$ exactly reduce to the PST tranformations in eq.~(\ref{PSTtrans}). 

Concluding: we intially had $2d$ gauge fields and $d$ abelian gauge symmetries. Imposing the gauge choice $\partial _\pp\,fA_== \partial _=f A_\pp$ eliminates half of the gauge fields but introduces one new degree of freedom $f$ leaving one unfixed gauge symmetry which appears as the PST gauge symmetry in the way outlined above. The PST symmetry acts as a shift on $f$ allowing it to be used to put $f=\tau$ which leads to the Tseytlin doubled formulation.

 \subsection{Equations of motion in the PST doubled formalism}
  
We can now see how the desired chirality constraints, eq.~\eqref{eq:Chirality},
follow as equations of motion in this approach.  For a single chiral boson a clear explanation of this was provided in \cite{Lechner:1998ga} and we adapt this to the doubled string taking into account the twisted nature of the constraints. 

The equations of motion that follows from a variation in ${\mathbb{X}}$ of the doubled action  \label{eq:Ldoubled} can be expressed as,
\begin{equation} 
 0 =  \partial_\pp\, \left( {\cal H} P_+ \nabla_=\IX - \frac{\partial_= f }{\partial_\pp\, f} P_- \nabla_\pp\, \IX \right)+ \partial_= \left( {\cal H} P_- \nabla_\pp\, \IX - \frac{\partial_\pp\, f }{\partial_= f} P_+ \nabla_=  \IX \right)  \ .  
\end{equation}
Introducing a one-form with components, 
\begin{equation}
v_\pp = \frac{\partial_\pp\,f}{\sqrt{\partial_\pp\,f\partial_=f}  }  \ , \quad v_= = \frac{\partial_= f}{\sqrt{\partial_\pp\,f\partial_=f}  } \ ,
\end{equation}
allows the equations of motion to be recast as,  
\begin{equation}
0 = d (v \Lambda  )  \ , \quad \Lambda =   v_\pp\, {\cal H} P_+ \nabla_= \mathbb{X} -   v_=  {\cal H} P_- \nabla_\pp\, \mathbb{X}  \ . 
\end{equation}
The homogenous solution $\Lambda  =0$ corresponds exactly, after making use of the chiral projectors $P_\pm$, to the chirality constraint,
\begin{equation}
 P_+ \nabla_= \mathbb{X} = 0 \ , \quad  P_- \nabla_\pp\, \mathbb{X} = 0 \ , 
\end{equation}
i.e. the covariant version of eq.~\eqref{eq:Chirality} that incorporates the connection.  There is also an inhomogeneous solution of the form $\Lambda_I =  \Gamma_I(f)   \sqrt{\partial_\pp\,f\partial_=f}$ since then $v \Lambda  =  df \Gamma (f)$ is trivially closed. However this is a pure gauge piece; under the residual gauge symmetry $\delta {\mathbb{X}} = \mathbb{T}(f)$ we have, 
\begin{equation}
\begin{aligned}
\delta \Lambda &= \frac{1}{\sqrt{\partial_\pp\,f\partial_=f}}\left( \partial_\pp\, f {\cal H} P_+ \partial_= \mathbb{T} -\partial_= f {\cal H} P_- \partial_\pp\, \mathbb{T}  \right)\\ 
&= \sqrt{\partial_\pp\,f\partial_=f}    \left( {\cal H} P_+ -  {\cal H} P_-   \right) \mathbb{T}'  =  \sqrt{\partial_\pp\,f\partial_=f}  \eta \mathbb{T}'   \ , 
\end{aligned}
\end{equation}
which is of the correct form to be gauged away with $\Gamma_I(f)= \eta \mathbb{T}' $.  

Performing the variation with respect to $f$ yields an equation of motion, 
\begin{equation}
\begin{aligned}
0 =&\,   \partial_\pp\, \left[ \frac{1}{\partial_= f} (P_+\nabla_= \mathbb{X})^T {\cal H} (P_+\nabla_= \mathbb{X}) - \frac{\partial_= f }{(\partial_\pp\, f)^2} (P_-\nabla_\pp\, \mathbb{X})^T {\cal H} (P_-\nabla_\pp\, \mathbb{X}) \right]\\
& \, -  \partial_= \, \left[ \frac{\partial_\pp\, f}{(\partial_= f)^2} (P_+\nabla_= \mathbb{X})^T {\cal H} (P_+\nabla_= \mathbb{X}) - \frac{1 }{\partial_\pp\, f} (P_-\nabla_\pp\, \mathbb{X})^T {\cal H} (P_-\nabla_\pp\, \mathbb{X}) \right] \ . 
\end{aligned}
\end{equation} 
Here the projectors $P_\pm$ come in handy to show that this equation can be recast as 
\begin{equation}
0 = d \left(v \frac{\Lambda^T\eta \Lambda}{\sqrt{\partial_\pp\,f\partial_=f}   } \right) \, , 
\end{equation} 
and hence follows as a consequence of the field equation for $\mathbb{X}$.  That this does not give rise to extra dynamical equations is a manifestation of the PST gauge symmetry.

\subsection{Gauge fixing and the dilaton} 
In our above derivations we  introduced a gauge fixing condition,
\begin{equation}
0 = \partial_\pp \, f  A_= - \partial_=f  \, A_\pp \ . 
\end{equation}
Let us consider how this should be done in a path integral.  We begin with the ill-defined, 
\begin{equation}
Z= \int [d\mathbb{X}] [d A_\pp\,] [d A_= ] e^{ - i \int {\cal L}[A_\pp\, , A_= , \mathbb{X}]  } \ , 
\end{equation}
and insert the gauge fixing condition and Jacobian,
\begin{equation}
Z=   \int [d\mathbb{X}] [d A_\pp\,] [d A_= ] \delta (\partial_\pp\, f \, A_= - \partial_= f \, A_\pp \, ) \det \left(\partial_\pp\, f \partial_= - \partial_=  f \partial_\pp \, \right)   e^{ - i \int {\cal L}[A_\pp\, , A_= , \mathbb{X}]  } \, . 
\end{equation}
At this stage the  function $f$ should not be considered dynamical but rather it  is a fixed background object that defines a gauge fixing.  The delta function restricts the path integral and since this is just an algebraic equation one can solve it by replacing $A_=$ with $A_\pp \frac{\partial_= f} {\partial_\pp f}$.   Hence,
\begin{equation}
\begin{aligned}
Z[f]&=   \int [d\mathbb{X}] [d A_\pp\,] [d A_= ] \frac{1}{\partial_\pp f } \delta \left( \, A_= - \frac{\partial_= f}{\partial_\pp f} \, A_\pp \, \right) \det \left(\partial_\pp\, f \partial_= - \partial_=  f \partial_\pp \, \right)   e^{ - i \int {\cal L}[A_\pp\, , A_= , \mathbb{X}]  } \\ 
&=  \int [d\mathbb{X}] [d A_\pp\,]  \frac{1}{\partial_\pp f } \det \left(\partial_\pp\, f \partial_= - \partial_=  f \partial_\pp \, \right)   e^{ - i \int {\cal L}[A_\pp \, \, , A_\pp \frac{\partial_= f}{\partial_\pp f} , \mathbb{X}]  }   \\
&=  \int [d\mathbb{X}] [d A \,]  [db] [d c]    e^{ - i \int {\cal L}[A_, \mathbb{X}; f] + {\cal L}_{gh}[b, c; f]  }   \ ,
 \end{aligned}
\end{equation}
in which we made the final change of variables $A_\pp = A \partial_\pp\, f$ and the ghost Lagrangian is given by
\begin{equation}
{\cal L}_{gh} = \partial_\pp\, f\,  b\, \partial_=c - \partial_= f\,  b\, \partial_\pp\, c \ .
\end{equation}
The   PST  symmetry, which extends to the ghost sector as,
\begin{equation}
\begin{aligned}
\delta f &=  \varepsilon \ ,  \\
 \delta \IX &=  \frac{ \varepsilon }{\partial _\pp\,f}P_-\, \nabla _\pp\,\IX+
\frac{ \varepsilon }{\partial _=\,f}P_+\, \nabla _=\,\IX\,, \\
\delta b &= \frac{1}{2} \varepsilon \left( \frac{\partial_\pp\, b}{\partial_\pp\,f}  - \frac{\partial_=  b}{\partial_= f} \right) \ ,  \\
\delta c &= \frac{1}{2} \varepsilon \left( \frac{\partial_\pp\,c}{\partial_\pp\,f}  - \frac{\partial_=  c}{\partial_= f} \right)  \ ,
\end{aligned}
\end{equation}
 can now be re-interpreted as saying nothing more than $Z[f]$ does not depend on the gauge fixing choice.  We can then simply choose to integrate over choices of the gauge fixing function $f$ in much the same way as one averages over gauge choices to obtain $R_\xi$ gauge in QED.  That is we can consider, 
 \begin{equation}
 Z = \frac{1}{vol_{PST} } \int [d f] Z[f] \ ,
 \end{equation}
 in which we divide by the volume of the PST group.    Since the PST  symmetry acts a simple shift on $f$, it can be fixed without the need for further ghost terms.

 To progress to the doubled formalism we now need to integrate out the gauge fields $A$ in this path integral.   As is well known, under T-duality, the dilaton receives a shift which in the Buscher procedure can be attributed to the determinant that comes from the Gaussian integral over the gauge fields.
 A useful mnemonic to obtain the correct shift is that the string frame supergravity measure $\sqrt{|g|}e^{-2 \phi}$ should be invariant.  For $g\rightarrow g^{-1}$ this means that T-dual dilaton is given by
\begin{equation}
\phi' = \phi - \frac{1}{2} \ln \det g \ . 
\end{equation}
On the other hand a T-duality invariant ``doubled dilaton'' is given by 
\begin{equation}\label{eq:doubledil}
\Phi = \phi - \frac{1}{4} \ln \det g  \ . 
\end{equation}

We can see that in the above derivation it is this doubled dilaton that emerges automatically in the covariant doubled formalism for elementary reasons; whereas in a tradition Buscher procedure on integrates out {\em two} components of a gauge field in the Gaussian term $ A_\pp\,  g A_=$ giving essentially a factor of $\det(g)^{-1}$, in the covariant fixing we have a Gaussian term $A g A \partial_\pp\, f \partial_= f$ and we integrate over a {\em single} mode, $A$, giving rise to a determinant factor $\det(g)^{-\frac{1}{2}} \times ( \partial_\pp\, f \partial_= f)^{-\frac{d}{2}} $.  The determinant of the metric enters with half the power and thus will give rise to a Fradkin Tseytlin coupling of to the doubled dilaton eq.~\eqref{eq:doubledil}.   Note that even if we begin with a non-flat geometry in which the normal dilaton is constant the doubled dilaton will not be.

\subsection{A comment on chiral gauging} 
\def\Em{\varepsilon^{=}}
\def\ep{\varepsilon^{\pp\,}}
\def\Dp{\partial_{\pp\,}}
\def\Dm{\partial_{=}}
\def\cH{{\cal H}}
\def\Pp{P_+}
\def\Pm{P_-}
\def\hpp{h_{\pp\, \pp\,}}
\def\hmm{h_{= =}}
In the derivation above we started with the usual string $\sigma$-model and performed an unusual gauge fixing in a Buscher procedure to obtain the manifestly Lorentz covariant doubled sigma model whose equations of motion imply chirality conditions.   One might wish to adopt a different tactic namely to begin with a {\em doubled} sigma-model from the outset and invoke the constraints via a gauging procedure.  In a previous paper,  \cite{Sevrin:2013nca}, two of the present authors emphasized that PST style actions for (supersymmetric) chiral bosons can be obtained by gauging a chiral (super)-conformal symmetry and by then specifying a Beltrami parametrisation for the corresponding gauge field.  This approach also works in the current case although in a rather surprising way which we will now illustrate (suppressing spectators for simplicity).   

We start with a Hull style $\sigma$-model on the doubled space,
   \begin{equation} \label{eq:Shull}
 	S_{Hull} = \frac{1}{2} \int d^2 \sigma \,    {\cal H}_{IJ}(y) \partial_\pp\, \IX^I   \partial_=\, \IX^J +\dots  \, ,
 \end{equation}
in which the ellipses indicate spectator terms that will play no role in what follows.   We want to  furnish the action with a gauge invariance, 
\begin{equation} 
\delta  \IX =   \varepsilon^\pp\,  P_-\, \partial_\pp\,\IX+ \varepsilon^=\,P_+\, \partial_=\,\IX\,,
\end{equation}
such that  only the field configurations obeying the constraint eq.~\eqref{eq:Chirality} are physical.   
{\it A priori} the gauge parameters $\Em$ and $\ep$   correspond to independent symmetries however as we shall soon see gauge invariance will force them to be related.  It is curious that in the ungauged action that this putative symmetry does not correspond to a  rigid invariance (unless $\partial_y {\cal H} = 0$); this is one of the features that makes the following gauging procedure rather atypical.   
We proceed by   introducing gauge fields $\hpp$ and $\hmm$ (not to be confused with the usual worldsheet metric components) with the usual conformal transformation rules,
   \begin{equation} 
 	  \begin{aligned}
	\delta \hpp &= \Dp\Em + \Em \Dm \hpp - \Dm \Em \hpp \ , \\ 
	\delta \hmm &= \Dm\ep + \ep \Dp \hmm - \Dp \ep \hmm \ ,
	  \end{aligned}
\end{equation}
and ``covariant'' derivatives, 
  \begin{equation} 
  \nabla_{\pp\,}^h \IX^{I} = \Dp \IX^{I} - \hpp (\Pp \Dm \IX)^{I} \ , \quad   \nabla_{=}^h \IX^{I} = \Dm \IX^{I} - \hmm (\Pm \Dp \IX)^{I}  \ . 
  \end{equation}
  In fact, though their structure is informed by the usual conformal covariant derivative, these derivates are not at all covariant as e.g. $\delta \nabla^h_= \IX  |_{\nabla^h_= \IX = 0 } \neq 0$.   That these derivatives are not actually covariant  makes the fact that the following construction works even more surprising.  We   continue regardless of this and consider the ``gauged''  action,
   \begin{equation} \label{eq:SgaugedChiral}
 	S_{gauged} = \frac{1}{2} \int d^2 \sigma \,    \cH_{IJ}  \, \nabla^h_{\pp\,} \IX^I \nabla^h_{= } \IX^J +   \dots \, . 
 \end{equation}
 Performing a gauge variation,   integrating by parts all terms containing $\partial_{\pp\, \pp\,} \IX$ and $\partial_{= =} \IX$ and making use of the identities obeyed by the projectors eq.~\eqref{eq:proj} results in a variation of the Lagrange density, 
\begin{equation}
  \begin{aligned}
\frac{1}{2} \delta {\cal L}_{gauged}   = & - \ep \Dp \cH_{{IJ}} (\Pm \Dp \IX)^{I} \Dm \IX^{J} + 2 \Em \hmm \eta_{IJ} (\Pm \Dp \Pp \Dm \IX)^{I} \Dp \IX^{J} \\ 
&- \Em \Dm \cH_{{IJ}}  \Dp \IX^{I} (\Pp \Dm \IX)^{J} - 2 \ep \hpp \eta_{IJ}   \Dm \IX^{I} (\Pp \Dm \Pm \Dp) \IX^{J} 
  \end{aligned}
  \end{equation}
 To cancel this we see that the gauge variation parameters are not independent and one must enforce, 
\begin{equation}\label{eq:gaugerelation}
 \hpp \hmm= 1 \ , \quad  \ep \hpp = \Em \ , \quad \Em \hmm = \ep \ . 
  \end{equation}
  It is easy to see that these are consistent with the gauge transformations rules.  With these identifications and the definitions of the projectors we find that indeed action eq.~\eqref{eq:SgaugedChiral}  is gauge invariant.   Solving the first of these relations with a Beltrami parametrisation
 \begin{equation}
  \hpp = \frac{\Dp f}{\Dm f} \ , \quad \hmm = \frac{\Dm f} {\Dp f} \ . 
 \end{equation}
and noting that the quadratic term in gauge fields vanishes by virtue of  $(\Pp)^{T} \cH \Pm = 0 $, one immediately recovers from   eq.~\eqref{eq:SgaugedChiral}  the Lorentz covariant action PST action of eq.~\eqref{eq:PST}.

 \section{Application to Non-Abelian T-duality}

Let us now consider the generalisation of these ideas to a non-Abelian group\footnote{In this work we restrict our attention to the cases in which the structure constants of the group dualised are traceless; this is to avoid the occurrence of a mixed gravitational-gauge anomaly when coupled to a curved background which upon dualisation can give rise to a Weyl anomaly i.e. a dual background that does not obey the (super)gravity equations. For discussion of this and related issues  see \cite{Giveon:1993ai,Alvarez:1994np,Elitzur:1994ri}.   }  of isometries, and for clarity we ignore spectator fields first and then give the result with their inclusion after.  Let us consider a  $\sigma$-model on a $d$-dimensional group space  $G$ specified by the Lagrange density, 
\begin{equation}\label{eq:nabtdual1}
{\cal L} =   L^a_\pp \, E_{ab} L^b_=  \ , 
\end{equation}
in which $E_{ab}$ is a constant (or possibly spectator dependant) matrix and the $L^a$ are the pull back to the worldsheet of the left invariant Maurer-Cartan forms for a group element $g \in G$ with conventions, 
\begin{equation}
L^a = - i \delta^{ab} \Tr T_b g^{-1} d g \ , \quad dL^a = \frac{1}{2} f^a{}_{bc} L^b \wedge L^c \ , \quad    [T_a , T_b] = i f_{ab}{}^c T_c  \ , \quad  \Tr T_a T_b = \delta_{ab} \, .
\end{equation}
This  $\sigma$-model has a global $G_L$ invariance that we can gauge by introducing a connection one-form $A= i A^a T_a$  in the algebra of  $G$ which minimally couples through the introduction of covariant derivatives,
\begin{equation}
\partial g \rightarrow Dg = \partial g - A g  \ . 
\end{equation}
The connection  has a field strength,
\begin{equation}
F_{\pp\,\, =} = \partial_\pp\, A_= - \partial_= A_\pp - [ A_\pp\, , A_= ] \ . 
\end{equation}
We see then that the gauging replaces the Maurer-Cartan forms with, 
\begin{equation}
L^a \rightarrow L^a - A^a D_{ab} \ , \quad D_{ab}= \Tr(T_a g T_b g^{-1}) \ ,
\end{equation}
in which we have defined the adjoint action $D_{ab}$ which obeys $D.D^T = {\bf 1}$. 
Then the action is invariant under the $G_L$ local transformations, 
\begin{equation}
g\rightarrow h^{-1}g \ , \quad  A \rightarrow h^{-1} A  h - h^{-1} \partial h \ . 
\end{equation}
In addition we introduce a Lagrange multiplier term $\Tr v F_{\pp\,\, = } $  to enforce a flat connection  which is gauge  invariant provided the Lagrange multipliers  transforms in the adjoint,
\begin{equation}
v \rightarrow h^{-1}v h \ . 
\end{equation} 
After integration by parts of the Lagrange multiplier term one finds a gauged   Lagrange density,
\begin{equation}
{\cal L}_{gauged} = L_\pp^T \,  E L_= - A^T_\pp \,  DE L_= - L_\pp^T E D^T A_= + A_\pp\, D E D^T A_= + A_\pp^T \partial_= v - A_=^T \partial_\pp\, v +  A_\pp \, F A_= \ ,
\end{equation} 
in which $F_{ab} = - i f_{ab}{}^c v_c$.  Obtaining the non-Abelian T-dual is then achieved by gauge fixing $g$ to the identity and integrating out the gauge fields to yield,  
\begin{equation}\label{eq:nabtdual2}
{\cal L}_{dual} = \partial_\pp \, v^T ( E +F )^{-1} \partial_= v \ . 
\end{equation}

Now we invoke the covariant gauge fixing choice, 
\begin{equation}
A_\pp^a = A^a \partial_\pp\, f \ , \quad A_=^a = A^a \partial_= f \ ,
\end{equation} 
and integrate out the field $A$.  Since the non-Abelian term in the field strength $[A_\pp \, , A_=]$ vanishes in this gauge the manipulations are actually quite similar to the Abelian case described earlier. 

If we define, 
\begin{equation}
\label{eq:Ldef}
\mathbb{L}^A  =  \left( \begin{array}{c}  L^a \\ \tilde{L}_a \end{array} \right)   \ , \quad  \tilde{L}_a = D_{ba}(g) \partial v^b \ , 
\end{equation}
then one finds a doubled action, 
\begin{equation}
\label{eq:NabTdouble}
{\cal L} =  \frac{1}{2} \mathbb{L}_\pp^T \, {\cal H}  \mathbb{L}_=  - \frac{1}{2}  \mathbb{L}_\pp^T  \, \Omega  \mathbb{L}_=  - \frac{1}{2} \frac{\partial_=  f}{\partial_\pp\, f}  \mathbb{L}_\pp^T \, ({\cal H}P_-) \mathbb{L}_\pp \,
  - \frac{1}{2} \frac{\partial_\pp\, f}{\partial_= f}  \mathbb{L}_=^T ({\cal H}P_+)  \mathbb{L}_= \ . 
\end{equation}

Notice that  the pull back of $\Omega_{AB}   \mathbb{L}^A \wedge   \mathbb{L}^B =  2 L^a \wedge \tilde{L}_a$  which entered the action as a purely topological term in the Abelian case is no-longer topological, instead it serves as a Kalb-Ramond potential. Since,  
\begin{equation}
d\tilde L_a = d(D_{ba} d v^b) = f_{ab}{}^c L^b \wedge \tilde{L}_c \ , 
\end{equation}
this implies a three-form  flux
\begin{equation}\label{eq:Hflux}
H= d(L^a \wedge \tilde L_{a} ) =- \frac{1}{2}f_{bc}{}^a L^b \wedge L^c \wedge \tilde{L}_a \ . 
\end{equation}

It is quite straightforward to extend these considerations to include a fibration and spectator coordinates.  Starting with the Lagrangian, 
\begin{eqnarray}
{\cal L}= L ^T_\pp\,  \,E \,L _= + L^T_\pp \,M \,\partial _=y+\partial _\pp\, y^T\,N \, L_= +
\partial _\pp\, y^T\,K \,\partial _=y\, ,
\end{eqnarray}
in which $E,M,N,K$ may have arbitrary dependence on the coordinates $y$, and repeating the above procedure yields the doubled action,
\begin{equation}
\begin{aligned}
 {\cal L} =&  \frac{1}{2} \mathbb{L}_\pp^{\nabla\, T} \, {\cal H}  \mathbb{L}^\nabla_=  - \frac{1}{2}  \mathbb{L}_\pp^{ T}   \, \Omega  \mathbb{L}_=  - \frac{1}{2} \frac{\partial_=  f}{\partial_\pp\, f}  \mathbb{L}_\pp^{\nabla\, T}  \, ({\cal H}P_-) \mathbb{L}^\nabla_\pp \,
  - \frac{1}{2} \frac{\partial_\pp\, f}{\partial_= f}  \mathbb{L}_=^{\nabla\, T}  ({\cal H}P_+)  \mathbb{L}^\nabla_=  \\ 
  & +\frac{1}{2} \mathbb{L}^T_\pp \eta \mathbb{B}_= - \frac{1}{2} \mathbb{B}_\pp^T \eta \mathbb{L}_=  +  \hat{K}_{\mu \nu}\partial_{\pp\,} y^\mu \partial_= y^\nu \ ,
  \end{aligned}
\end{equation}
in which we defined,
\begin{equation}
 \mathbb{L}^\nabla = \mathbb{L} + \mathbb{B} \ , \quad   P_+ \mathbb{B}  = \left(\begin{array}{cc} g^{-1} M \\ E^T g^{-1} M   \end{array} \right) \ , \quad P_- \mathbb{B}  = \left(\begin{array}{cc} g^{-1} N^T \\  - E g^{-1} N^T   \end{array} \right) \, , 
\end{equation}
and the modified Lagrangian on the base involves $\hat{K}$ defined as in the Abelian case in eq.~\eqref{eq:baseK}.

\subsection{Relation to Poisson Lie Doubled Formalism}
 
There is  an existing formulation for a non-Abelian T-duality double formalism, which in fact also accommodates a  further generalisation known as Poisson Lie T-duality \cite{Klimcik:1995ux,Klimcik:1995dy}.   The result we obtained  in eq. \eqref{eq:NabTdouble} can be understood in this context.  To do so we remind the reader of a little technology -- the Drinfeld double \cite{Drinfeld:1986in}.

  The Drinfeld double ${\cal D}$ is a Lie algebra that can be decomposed as the sum of two sub algebras  ${\cal D}= {\cal G} \oplus \tilde{\cal G}$ that are  maximally isotropic with respect to an inner product $\langle \cdot | \cdot \rangle$.  If $T_a$ are the generators of  ${\cal G}$ and $\tilde{T}^a$ those of   $\tilde{\cal G}$,  then the generators of the double $\mathbb{T}_A = \{ T_a , \tilde T^a\}$ obey   $\eta_{AB} =  \langle \mathbb{T}_A  | \mathbb{T}_B \rangle$ i.e.,
\begin{equation}
  \langle T_a  | T_b \rangle =   \langle \tilde T^a  | \tilde T^b \rangle = 0 \ , \quad \langle T_a | \tilde T^b \rangle = \delta_a{}^b\ .
\end{equation}
  The structure constants of the double $[\mathbb{T}_A,\mathbb{T}_B] = i F_{AB}{}^C \mathbb{T}_C$ decompose as, 
\begin{equation}
[T_a , T_b] = i f_{ab}^c T_c \ , \quad [\tilde T^a , \tilde T^b ] = i \tilde{f}^{ab}{}_c \ , \quad [T_a , \tilde{T}^b ]  = i \tilde{f}^{bc}{}_a T_c - i f_{ac}{}^b\tilde{T}^c  \ , 
\end{equation}
and the Jacobi identity places further constraints on the admissible choices of ${\cal G}$ and ${\cal \tilde{G}}$.   We also need to define some matrices for $g \in G$ the group of ${\cal G}$, 
\begin{equation}
g^{-1} T_a g = a_a{}^b T_b \ , \quad g^{-1} \tilde{T}^a g = b^{ab} T_b + (a^{-1})_{b}{}^a \tilde{T}^b \ , \quad \Pi^{ab}= b^{ca}a_{c}{}^b \ ,
\end{equation}
and tilde analogues, $\tilde{a},\tilde{b},\tilde{\Pi}$, for $\tilde{g} \in \tilde{G}$.   The statement of Poisson-Lie T-duality then is the equivalence between the two $\sigma$-models,
\begin{equation}
\begin{aligned}\label{eq:PLpairs}
S= \int d^2 \sigma (E^{-1} + \Pi)^{-1}_{ab} L_\pp\,^a L_=^b \ , \quad \tilde{S} = \int d^2\sigma [(E+ \tilde \Pi)^{-1}]^{ab}\check{L}_{\pp\, a} \check{L}_{= \, b} \ , 
\end{aligned} 
\end{equation}
where $L$ and $\check{L}$ refer to the left-invariant one-forms of $G$ and $\tilde{G}$ respectively (a h\'a\v{c}ek is used to distinguish   $\check{L}$ from $\tilde{L}$ introduced above).   

If $G= \tilde{G}= U(1)^d$ we have an Abelian double and the dual pairs of $\sigma$-models correspond to Abelian  T-duals.  If  ${\cal G}$ is the algebra of some $d$-dimensional non-Abelian Lie group and  $\tilde{\cal G}= u(1)^d$, the double is said to be semi-Abelian and  the two dual models in eq.~\eqref{eq:PLpairs} reduce exactly to non-Abelian T-dual related actions of eq.~\eqref{eq:nabtdual1} and eq.~\eqref{eq:nabtdual2}.   The case where neither $\tilde{G}$ nor $G$ are Abelian corresponds to a dualisation of non-isometric 
$\sigma$-models  and has recently found new applications in the context of the relation between certain classes of integrable models in two dimensions known as $\eta$ and $\lambda$ deformations\footnote{For a brief summary of  this direction   the reader may consult \cite{Thompson:2015lzd} and references within.}.

That the actions in eq.~\eqref{eq:PLpairs} are dual was established in  \cite{Klimcik:1995ux,Klimcik:1995dy} by constructing an action on the Drinfeld double given by,
\begin{equation}\label{eq:PLdouble}
S_{PLT} = \frac{1}{2}\int_\Sigma \langle l^{-1}\partial_\sigma l |  l^{-1}\partial_\tau l \rangle   +\frac{1}{12}\int_{M_3} \langle l^{-1} dl | [l^{-1} dl , l^{-1} dl ]\rangle - \frac{1}{2}\int_\Sigma  \langle l^{-1}\partial_\sigma l|{\cal H} |   l^{-1}\partial_\sigma l \rangle  \, ,
\end{equation}
in which $l$ is group element of the Drinfeld double, ${\cal H}_{AB} =  \langle \mathbb{T}_A | {\cal H}  | \mathbb{T}_B \rangle$ is just the $O(d,d)$ coset generalised metric  and $M_3$ is a suitable three-manifold whose boundary is the worldsheet $\Sigma$.  This action can be thought of as deforming the chiral WZW model of Sonnenschein \cite{Sonnenschein:1988ug} and is essentially a doubled action in a Tseytlin style non-covariant gauge.   Parametrising $l = \tilde{g} g$ with $\tilde g \in \tilde G$ and $g \in G$ and integrating out $\tilde{g}$ will give the action $S$ of eq.~\eqref{eq:PLpairs} and doing the converse with $l = g \tilde{g}$ gives the dual action $\tilde{S}$.

There also exists a PST version of the doubled action eq.~\eqref{eq:PLdouble} given by \cite{KSnotes},\footnote{To the best of our knowledge this has not appeared in the literature    and we are grateful to K. Sfetsos for sharing his notes in which it was derived.}
\begin{equation}
\begin{aligned}\label{eq:PLT-PST}
S_{PLT-PST}  =&  \frac{1}{2}\int_\Sigma \langle l^{-1}\partial_\pp\, l | {\cal H} |  l^{-1}\partial_= l \rangle +\frac{1}{12}\int_{M_3} \langle l^{-1} dl | [l^{-1} dl , l^{-1} dl ] \rangle   \\
&  - \frac{1}{2}\int_\Sigma \frac{\partial_\pp\, f}{\partial_= f }    \langle l^{-1}\partial_= l | {\cal H} P_+ |  l^{-1}\partial_= l \rangle +  \frac{1}{2}\int_\Sigma \frac{\partial_= \, f}{\partial_\pp\, f }    \langle l^{-1}\partial_\pp\, l | {\cal H} P_- |  l^{-1}\partial_\pp\, l \rangle \, .
 \end{aligned}
\end{equation}

Let us now restrict our attention to the semi-Abelian double appropriate for non-Abelian T-duality.  The first thing to note is that if we express the group element on the double as $l = \tilde{g} g$ then,
\begin{equation}
l^{-1} dl = g^{-1} \tilde{g}^{-1} d\tilde{g} g + g^{-1} d g = i d v_a g^{-1} \tilde{T}^a g + g^{-1} d g = i a^{-1}(g)_{b}^a dv_a \tilde{T}^b + i L^a T_a \, ,
\end{equation} 
in which we parametrised $\tilde{g}= \exp( i v_a \tilde{T}^a)$.  However since $a^{-1}(g)$ is no more than  the adjoint action, $D^T(g)$, we see that,
\begin{equation}
l^{-1} d l = i \tilde{L}_a \tilde{T}^a + i L^a T_a = i {\mathbb L}^A {\mathbb T}_A 
\end{equation} 
coinciding with the definition in eq.~\eqref{eq:Ldef}.   One can now see that all the terms involving ${\cal H}$ in \eqref{eq:NabTdouble} directly match those in eq.~\eqref{eq:PLT-PST}. All that remains is to understand the WZ term for which we observe, 
 \begin{equation}
\langle l^{-1} dl | [l^{-1} dl , l^{-1} dl] \rangle = F_{AB}{}^C \eta_{CD}  {\mathbb L}^A\wedge {\mathbb L}^B  \wedge {\mathbb L}^C = 3 f_{ab}{}^c L^a \wedge L^b \wedge \tilde{L}_c 
\end{equation} 
which is in agreement with eq.~\eqref{eq:Hflux}, thus confirming what started off as a topological term in the Abelian doubled theory is precisely what is needed as a potential for the WZ in the non-Abelian doubled theory. 

To close this section let us finally note that actions of this style  have been used in 
  \cite{Hull:2007jy,Hull:2009sg,Dall'Agata:2008qz,Avramis:2009xi}    to describe strings whose doubled target space is a twisted torus and have been conjectured to give a world sheet description of ${\cal N}=4$ electrically gauged supergravities.  The works   \cite{Hull:2007jy,Hull:2009sg} have the chirality constraint as supplementary to the action and those of  \cite{Dall'Agata:2008qz,Avramis:2009xi} use the Tseytlin style formulation.   It will be of interest to make more precise the linkage between the spacetime violation of section condition leading to gauged supergravities as in  \cite{Geissbuhler:2011mx,Aldazabal:2011nj,Grana:2012rr,Berman:2012uy} and the generalised notions of Poisson-Lie duality whose worldsheet generalised metric has dependence on both coordinates and their duals.

  \section{Towards the Supersymmetric Doubled String}
A supersymmetric first order manifest T-dual invariant worldsheet formulation is still lacking. Even a non-covariant Tseytlin type description has not been given yet. We provide here a first step by constructing the simplest model which has an ${\cal N}=(0,1)$ worldsheet supersymmetry. While extremely simple it already exhibits all subtleties which also occur in models with more supersymmetry. We will keep supersymmetry manifest by working in ${\cal N}=(0,1)$ superspace (conventions can be found at the beginning of appendix A).

\subsection{The covariant formulation}
For simplicity we restrict ourselves to a trivial bundle structure. All results can rather straightforwardly be generalized to a non-trivial bundle structure. The starting point is the Lagrange density,
\begin{eqnarray}
{\cal L}=2i\, \partial _\pp\, x\,E\,D_-x+ {\cal L}_S(y)\,,
\end{eqnarray}
where $x$ is a set of adapted coordinates such that the background field $E=E(y)$ depends only on the spectator coordinates $y$ whose dynamics is governed by $ {\cal L}_S$. In order to gauge the isometries $x \rightarrow x+ \varepsilon $ we introduce gauge fields $A_\pp$ and $A_-$ and using Lagrange multipliers $\tilde x$ we impose flatness. The gauged $ \sigma $-model is given by\footnote{Note that we could as well have introduced the full ${\cal N}=(0,1)$ gauge multiplet which consists of $A_\pp\,$, $A_=$ and $A_-$. Introducing Lagrange multipliers which constrain all fieldstrengths $F_{\pp\,=}$, $F_{\pp\,-}$, $F_{=-}$ and $F_{--}$ to zero, one finds that upon making a field redefinition on $\tilde x$  this  reduces to the current case.},
\begin{eqnarray}
{\cal L}=2i\, \partial _\pp\, x\,E\,D_-x+2i\,A_\pp\,E\,A_- +2i\,A_\pp\,{\cal J}_-+ 2i\,{\cal J}_\pp\, A_-+
{\cal L}_S(y)\,,
\end{eqnarray}
where,
\begin{eqnarray}
{\cal J}_\pp\,= E^T \partial _\pp\,x - \partial _\pp\,\tilde x\,,\qquad
{\cal J}_-=E\,D_-x+D_-\tilde x\,.
\end{eqnarray}
Integrating over $\tilde x$ gives the original model back. Motivated by the non-supersymmetric case we impose the gauge choice,
\begin{eqnarray}
A_\pp\,= \partial _\pp\,f\,A\,,\qquad
A_-= D_-f\,A\,,
\end{eqnarray}
where $f$ is an arbitrary function and $A$ is a set of $d$ ${\cal N}=(0,1)$ scalar superfields. The Lagrange density becomes so,
\begin{eqnarray}
{\cal L}=2i\, \partial _\pp\, x\,E\,D_-x+2i\, \partial _\pp\,fD_-f A\,g\,A +2i\,A\,\big(\partial _\pp\,f\,{\cal J}_-+ 
D_-f\,{\cal J}_\pp\,\big)+
{\cal L}_S(y)\,.\label{ld1o}
\end{eqnarray}
The residual gauge invariance is given by,
\begin{eqnarray}
x \rightarrow x+ \varepsilon (f)\,,\qquad A \rightarrow A-  \partial _f \varepsilon(f) \,, \qquad
f \rightarrow  f\,.
\end{eqnarray}
In addition the action is invariant under,
\begin{eqnarray}
\tilde x \rightarrow  \tilde x+ \tilde \varepsilon (f)\,,
\end{eqnarray}
as well.
The equations of motion for $A$ are given by,
\begin{eqnarray}
D_-f\,A=-\frac 1 2 \,g^{-1} {\cal J}_-- \frac 1 2 \, \frac{D_-f}{ \partial _\pp\,f}\,g^{-1}J_\pp\,,\label{eom01}
\end{eqnarray}
which, because of the fermionic nature of $D_-f$, {\em cannot} directly be solved for $A$. However one notes that by 
multiplying the equations of motion by $D_-f$ one obtains the constraint,
\begin{eqnarray}
D_-f\, {\cal J}_-=0\,.\label{nilcon}
\end{eqnarray}
Acting with $D_-$ on this one gets,
\begin{eqnarray}
{\cal J}_-=2i\, \frac{D_- f}{ \partial _=f }\,D_- {\cal J}_-\,.\label{nilcon1}
\end{eqnarray}
Despite appearances, eq.~(\ref{nilcon}) is an algebraic constraint on the components of the superfields. Indeed writing the superspace components of $x$ and $\tilde x$, $x=x +i \theta^-\psi_-$ and $\tilde x=\tilde x +i \theta^-\tilde\psi_-$, one readily verifies using eq.~(\ref{nilcon1}) that  the constraint can be solved for half of the component fields $\psi$ and $\tilde \psi$. As such this constraint can simply be imposed using Lagrange multipliers. This is very reminiscent of the nilpotent superfield constraints \cite{Rocek:1978nb}. 
Using this in the equations of motion eq~(\ref{eom01}) one solves for $A$,
\begin{eqnarray}
A=- \frac 1 2 \, \frac{1}{ \partial _\pp\,f}\,g^{-1}{\cal J}_\pp-\frac{i}{ \partial _=f} \,g^{-1} D_-{\cal J}_-+D_-f\,\big(\cdots\big)\,,
\label{eom02}
\end{eqnarray}
where the terms following $D_-f$ remain undetermined but they will not play any role in what follows. Using this to eliminate $A$ in the first order Lagrange density eq.~(\ref{ld1o}) one gets,
\begin{eqnarray}
{\cal L }&=&2i\, \partial _\pp\, x\,E\,D_-x
-i \, {\cal J}_\pp\,g^{-1}\, {\cal J}_-
- \frac i 2 \, \frac{D_-f}{ \partial _\pp\,f}\, {\cal J}_\pp\, g^{-1}\, {\cal J}_\pp\, \nonumber\\
&&+ \frac{\partial _\pp\,f }{ \partial _=f}\, {\cal J}_-\,g^{-1}\,D_- {\cal J}_-
+{\cal L}_S(y)\,,\label{ld2o}
\end{eqnarray}
together with the constraint given in eq.~(\ref{nilcon}).
Repeatedly using eqs.~(\ref{nilcon}) and (\ref{nilcon1}), one rewrites this as,
\begin{eqnarray}
{\cal L}&=&2\, \frac{ \partial _\pp\,f}{ \partial _=f}\Big(D_-- \frac{D_-f}{ \partial _\pp\,f}\,\partial _\pp\,\Big)\IX
\,\eta\, P_+D_-\big(P_+D_-\IX\big)-i\,
\Big(D_-- \frac{D_-f}{ \partial _\pp\,f}\,\partial _\pp\,\Big)\IX
\,\eta\, P_- \partial _\pp\,\IX\, \nonumber\\
&& +2i\, \frac{D_-f}{ \partial _=f}\,  \Psi_{+}\,\eta\,P_+\,D_-\IX+{\cal L}_S(y)\,,\label{n01covac}
\end{eqnarray}
where the topological term has been dropped. The Lagrange multiplier $\Psi_+$, which transforms under $O(d,d;\IZ)$ in the same way as $\IX$, enforces the constraint,
\begin{eqnarray}\label{eq:constraint}
\mu_{=} \equiv D_-f\,P_+\,D_-\IX=0\,,
\end{eqnarray} 
which is equivalent to the constraint in eq.~(\ref{nilcon}). Note that because of the presence of the projection operator $P_+$  only $d$ components of the Lagrange multiplier $\Psi_+$ effectively appear in the lagrangian. 

The covariant action has two classes of symmetries,
\begin{description}
\item [The residual gauge symmetry] Even after we gauge fixed the gauged non-linear $\sigma $-model there is a residual gauge invariance left: 
\begin{eqnarray}
 f& \rightarrow  & f \nonumber\\
  \IX& \rightarrow  & \IX+ \Lambda(f)\nonumber\\
 \Psi_+ & \rightarrow  & \Psi_+-\frac i 2 \,D_-S\, \partial _\pp \,\Lambda \,,\label{ressyma}
\end{eqnarray}
where $\Lambda(f)$ is a $2d\times 1$ column matrix of arbitrary functions of $f$.
\item [The PST symmetry]
As the  gauge fixing function $f$ was randomly chosen, we expect that it can be shifted in an arbitrary way which is the origin of the PST symmetry.  To see this first consider the action defined by eq.~\eqref{n01covac} in the absence of the Lagrange multiplier term.  After some significant effort one determines that under the variation,
\begin{eqnarray}
\delta f&=& \varepsilon   \\
\delta \IX&=&  \frac{\varepsilon}{\partial_{\pp\,} f } P_- \partial_{\pp\,} \IX + \frac{2 i \varepsilon}{\partial_{=} f } P_+ D_- \left( P_+ D_- \IX \right) - \frac{2 i \varepsilon}{\partial_=f} D_-f P_+ D_-\left(\frac{1}{\partial_{\pp\,} f } P_- \partial_{\pp\,} \IX  \right)  \nonumber \ ,
\end{eqnarray}
one produces only terms that are proportional to the constraint $\mu$ defined in eq.~\eqref{eq:constraint} or derivatives thereof.   Moreover, this property is shared by the variation of the constraint itself.   As a result, one is then guaranteed a transformation of the Lagrange multiplier that renders the whole Lagrangian \eqref{n01covac} invariant.   For pedagogical purpose we illustrate this in the simplest case of constant background fields in the Appendix.  

\end{description}

\subsection{The Tseytlin formulation}
We now pass to a Lorentz non-covariant gauge for the PST symmetry in order to recover a Tseytlin like formulation. Choosing $f=f(\tau)$ \cite{Sevrin:2013nca} we get that the Lagrange density eq.~(\ref{n01covac}) becomes,
\begin{eqnarray}
{\cal L}&=&-\frac i 2 \,\hat D\IX\,\eta\, \partial _\tau\IX+\frac i 2 \hat D\IX\, {\cal H} \,\partial_ \sigma \IX+
\hat D \IX\,\eta\,P_+D_-SD_-\IX+\nonumber\\
&& \theta^-\Psi_+\,\eta\,P_+\,D_-\IX+{\cal L}_S(y)\,, 
\end{eqnarray}
where,
\begin{eqnarray}
\hat D \equiv D_- +\frac i 2 \,  \theta^- \partial _\pp\,= \partial _-+\frac i 2 \, \theta^-  \partial _ \sigma \,,\qquad 
\hat D {}^2=\frac i 2 \, \partial _ \sigma\,,\qquad \theta^-\,D_-=\theta^-\,\hat D\,. 
\end{eqnarray}
The equations of motion for $\Psi_+$ and $\IX$ read,
\begin{eqnarray}
&&\theta^-P_+D_-\IX=\theta^-P_+\hat D\IX=0\,, \nonumber\\
&&\hat D\Big(\partial _ \tau\IX- S\, \partial _ \sigma  \IX +i\, \hat D S\, \hat D \IX+iP_+D_-S\,D_-\IX \nonumber\\
&&\qquad -i D_- S\, P_+\hat D \IX +\theta^-P_+\Psi_+\Big)=0\,.\label{ufid0}
\end{eqnarray}
The second of these equations  immediately implies,
\begin{eqnarray}
\partial _ \tau\IX- S\, \partial _ \sigma  \IX +i\, \hat D S\, \hat D \IX+iP_+D_-S\,D_-\IX 
 -i D_- S\, P_+\hat D \IX +\theta^-P_+\Psi_+=F(\tau)\,,\label{ufida}
\end{eqnarray}
with $F(\tau)$ an arbitrary function of $\tau$. Using the residual gauge invariance, eq.~(\ref{ressyma}), which assumes now the form,
\begin{eqnarray}
   \IX& \rightarrow  & \IX+ \Lambda(\tau)\nonumber\\
 \Psi_+ & \rightarrow  & \Psi_+-\frac i 4 \,D_-S\, \partial _\tau \,\Lambda(\tau) \,,\label{ressymb}
\end{eqnarray}
this function can be put to zero leaving us with,
\begin{eqnarray}
\partial _ \tau\IX- S\, \partial _ \sigma  \IX +i\, \hat D S\, \hat D \IX+iP_+D_-S\,D_-\IX 
 -i D_- S\, P_+\hat D \IX +\theta^-P_+\Psi_+=0\,.\label{ufidb}
\end{eqnarray}

The first equation in eq.~(\ref{ufid0}) implies,
\begin{eqnarray}
P_+\hat D\IX= \theta^-P_+\hat D\big(P_+\hat D \IX\big)\,.\label{ufid1}
\end{eqnarray}
Acting with $P_-$ on  eq.~(\ref{ufidb}) and using eq.~(\ref{ufid1}) one obtains,
\begin{eqnarray}
P_-\, \partial _\pp\,\IX=0\,.\label{con011}
\end{eqnarray}
Acting with $P_+$ on eq.~(\ref{ufidb}) allows one to solve for for $\theta^-P_+\Psi_+$. However multiplying this equation with $\theta^-$ gives,
\begin{eqnarray}
\theta^-P_+D_-\big(P_+D_-\IX\big)=0\,.\label{ufid2}
\end{eqnarray}
From the first equation in eq.~(\ref{ufid0}) one also gets,
\begin{eqnarray}
P_+ D_-\IX= \theta^-P_+ D_-\big(P_+ D_- \IX\big)\,
\end{eqnarray}
which combined with eq.~(\ref{ufid2}) gives,
\begin{eqnarray}
P_+ D_-\IX=0\,.\label{con012}
\end{eqnarray}
So the equations of motion of the model in the Tseytlin gauge indeed reproduce the constraints eqs.~(\ref{con011}) and (\ref{con012}) as expected.

\subsection{Component Form}
For  convenience we now give the results expanded into components as defined by the superfield expansion $\IX = \IX +i\, \theta^- \Xi_-$.  We use that $D_- \hat{D}|_{\theta^-=0} = \frac{i}{2} \partial_\sigma$ to find that the lagrangian can be expressed as, 
\begin{equation}
\begin{aligned}
D_- {\cal L}|_{\theta=0} = & \frac{1}{4} \partial_\sigma \IX \eta \partial_\tau \IX - \frac{1}{4} \partial_\sigma \IX {\cal H} \partial_\sigma \IX   \\ 
& -  i \,\Xi_-\, \eta \left( \partial_\pp\, \Xi_-\, +i\, D_-S\, \partial_\sigma \IX \right) + \frac{1}{2}\, \Xi_-\, \eta D_-S D_-S \Xi_-\, + i\,\Psi_+ \eta P_+\Xi_-\,  \ . 
\end{aligned}
\end{equation}
Here, and in the following component expressions, we adopt the implicit notation that $D_-S \equiv D_-S|_{\theta=0}$ and $\IX \equiv \IX|_{\theta=0}$.
Note the presence of a four-fermi interaction term that would have been hard to guess from the bosonic case; this term will prove essential in what follows. 
As above, the variation with respect to the Lagrange multiplier enforces,
\begin{equation}
\label{eq:componenteqm1}
 P_+ \Xi_- =0 \, . 
 \end{equation} 
The variation with respect to $\IX$ gives an equation of motion that is a total $\partial_\sigma$ derivative which,  using the residual gauge redundancy,  can be integrated to yield,
 \begin{equation}
 \label{eq:componenteqm2}
 P_+ \left( \partial_= \IX - D_-S\, \Xi_-\,  \right) =0 \ , \quad P_- \partial_\pp\,\IX = 0  \  . 
 \end{equation}
The variation with respect to the fermion is more intricate and yields,
\begin{equation}
2  \partial_\pp\,  \Xi_- + i D_-S \partial_\sigma \IX +i D_-S D_-S \,\Xi_-  + P_+ \Psi_+  = 0  \ . 
\end{equation} 
The $P_+$ projection of this equation fixes the Lagrange multiplier however the $P_-$ projection provides a fermion equation of motion, 
\begin{equation} 
\begin{aligned}
0 &= P_- \left(  \partial_\pp\,  \Xi_- + \frac{i}{2}\, D_-S \partial_\sigma \IX + \frac{i}{2}\, D_-S P_+ D_-S \,\Xi_-   \right)   \\
&=  P_- \left(  \partial_\pp\,  \Xi_- + \frac{i}{2}\, D_-S \partial_\sigma \IX  + \frac{ i}{2} D_-S \partial_= \IX   \right)   \\
&= P_-  \left(  \partial_\pp\,  \Xi_- + \frac{i}{2}\, D_-S \partial_\pp\, \IX    \right)    \ , 
 \label{eq:componenteqm3}
\end{aligned}
\end{equation}
in which we used that $P_- D_-S  = D_-S P_+$ and the equation of motion eq.~\eqref{eq:componenteqm2} to pass to the final line.  Together the equations~\eqref{eq:componenteqm1}, \eqref{eq:componenteqm2}, and \eqref{eq:componenteqm3} are exactly the component content of the superspace equations, 
\begin{equation}
P_+  D_- \IX = P_- \partial_\pp\, \IX = 0 \ . 
\end{equation}
\section{Discussion and open problems} 
In this paper we have clarified many missing details in the construct of the manifestly T-duality symmetric worldsheet theory and shown how such a formulation can be obtained through a novel gauge fixing choice.    This procedure allowed us to make the generalisation to the supersymmetric case in the most minimal, but still non-trivial, extension to ${\cal N}=(0,1)$ supersymmetry.   The essential reason for the complexity comes from having in the theory chiral bosons whose chirality is mis-aligned with that of the supersymmetry. 

The natural next direction here is to extend this work to both ${\cal N}=(1,1)$ and ${\cal N}=(2,2)$ supersymmetry.  The ${\cal N}=(1,1)$ case is already under study and will directly follow from the techniques outlined within.  The ${\cal N}=(2,2)$ remains less obvious but should be an exciting arena to make a direct link to Hitchin's generalised geometry.  Initial results in this direction have recently been reported by one of us \cite{AS}.   It will also be of interest to consider spacetime supersymmetry generalising the result of  \cite{Bandos:2015cha} to curved backgrounds. 

Our discussion has been local in nature and there are sensitive   issues, even in the bosonic theory, that will have to be addressed if the derivation used is to be implemented in full Polyakov sum over genus at the quantum level.  At first sight our gauge fixing  choice $\partial _\pp\,fA_= = \partial _=fA_\pp\,$ looks to require the introduction of a globally defined exact form $u=df$.  In fact this is too strong, as is known from previous studies of the PST formalism    \cite{Berman:1998va} it is sufficient to work with a closed form $du=0$. Put another way  \cite{Bandos:2014bva}, the residual gauge invariance is sufficient to eliminate cohomological contributions that come from integrating the equation of motion to produce the constraint.     However one still requires in the manipulations that $f$ has nowhere vanishing first derivatives  so as to allow such terms to appear in the denominator of fractions in a PST approach.    Since this necessitates a the existence of a nowhere vanishing vector field, it is not obvious how to extend from $\mathbb{R}^2$ to a compact Riemann surfaces of non-vanishing Euler character.   The appearance of the function $f$ was via a gauge fixing, the interpretation here is that the gauge fixing choice adopted can not be globally extended and is only locally well defined.  A possible resolution is to find a suitable global fixing or to work patchwise.  Understanding this will be an interesting topic for further investigation

This formulation may have great utility; by calculating the $\beta$-functions in a perhaps naive manner one could hope to find background field equations for the generalised metric which relate to the target space formulation of {\bf DFT}.  Whilst the non-covariant Tseytlin style action allows for such progress to be made at 1-loop order     \cite{Berman:2007xn,Berman:2007yf,Copland:2011wx}, it is very hard to extend this to higher loops -- the non-Lorentz invariant structure makes the regularisation of Feynman diagrams taxing at best.   Using the covariant formulation may alleviate some of this trouble.  Optimistically we hope that the techniques in this paper could prove to be a valuable starting point for the calculation of duality covariant corrections to {\bf DFT}.

\section*{Acknowledgments}
This work is supported in part by the Belgian Federal Science Policy Office through the Interuniversity Attraction Pole P7/37, and in part by the ``FWO-Vlaanderen'' through the project G020714N and a postdoctoral fellowship, and by the Vrije Universiteit Brussel through the Strategic Research Program ``High-Energy Physics''. We are grateful to J.P. Ang, David Berman, Chris Blair and Martin Ro\v cek and  for numerous illuminating discussions, and to K. Sfetsos for discussions and sharing with us his work on the PST formulation of Poisson-Lie duality.   We would like to thank the Simons Center for Geometry and Physics for providing a stimulating environment during the conference ``Generalized Geometry and T-dualities'' while part of this work was finalized.

\begin{appendix}
\section{Conventions}
Throughout the paper we use worldsheet lightcone coordinates,
\begin{eqnarray}
\sigma ^\pp= \tau + \sigma ,\qquad \sigma ^== \tau - \sigma\, .\label{App1}
\end{eqnarray}
In $N=(0,1)$ superspace this is extended by adding one one-component real fermionic coordinate $ \theta ^-$. The fermionic derivative $D_-$ satisfies,
\begin{eqnarray}
D_-^2=- \frac{i}{2}\, \partial _= \,.
\label{App2}
\end{eqnarray}

The T-duality group $O(d,d;\IZ)$ plays a central role. In the present context ${\cal O}\in O(d,d;\IZ)$ is a $2d\times 2d$ matrix with integer entries satisfying,
\begin{eqnarray}
{\cal O}^T\,\eta\, {\cal O}=\eta\,,
\end{eqnarray}
where,
\begin{eqnarray}
\eta= \left(
\begin{array}{cc}
0&{\bf 1}\\
{\bf 1}&0\end{array}
\right)\,.
\end{eqnarray}

In the current paper we use adapted coordinates $x^i$ and their T-duals $\tilde x_i$, $i\in\{1,\cdots , d\}$ together with spectator coordinates $y^\mu $, $\mu\in\{1,\cdots D-d\}$. We write the adapted coordinates together with the dual ones into a single $O(d,d;\IZ)$ multiplet,
\begin{eqnarray}
\IX=
\left(
\begin{array}{c}
x\\
\tilde x\end{array}
\right)\,, 
\end{eqnarray}
which transforms under the action of ${\cal O}\in O(d,d;\IZ)$ as,
\begin{eqnarray}
\IX \rightarrow \IX'= {\cal O}^{-1}\,\IX.
\end{eqnarray}
Writing ${\cal O}\in O(d,d;\IZ)$ as,
\begin{eqnarray}
{\cal O}=
\left(
\begin{array}{cc}
A&B\\
C&D\end{array}
\right)\, ,
\end{eqnarray}
the background fields $E_{ij}(y)=g_{ij}(y)+b_{ij}(y)$ transform non-linearly,
\begin{eqnarray}
E \rightarrow E'=(EB+D)^{-1}(EA+C)\,,
\end{eqnarray}
however, the {\em generalised metric} ${\cal H}$,
\begin{eqnarray}
{\cal H}=\left(
\begin{array}{cc}
1 & -b\\
0 & \,1 \end{array}
\right) \left(
\begin{array}{cc}
g &   \,\,0\\
0 &  \,g^{-1} \end{array}
\right)
\left(
\begin{array}{cc}
1 & 0\\
b & 1 \end{array}
\right)= 
\left(
\begin{array}{cc}
g-b\,g^{-1}\,b&\,-b\,g^{-1}\\
g^{-1}\,b&g^{-1}\end{array}
\right)\,,
\end{eqnarray}
 transforms linearly,
\begin{eqnarray}
{\cal H}\rightarrow {\cal H}'= {\cal O}^T {\cal H} {\cal O }\,.
\end{eqnarray}
From ${\cal H}$ we construct an almost product structure $S$,
\begin{eqnarray}
S=\eta\, {\cal H}\,,
\end{eqnarray}
such that $S^2=+{\bf 1}$. Using this we introduce the orthogonal projection operators $P_+$ and $P_-$,
\begin{eqnarray}
P_\pm\,=\,\frac 1 2 \,\big({\bf 1}\pm S\big)\,.
\end{eqnarray}
Some often used identities include,
\begin{eqnarray}\label{eq:proj}
\begin{aligned}
{\cal H}\,P_\pm = P_\pm^T {\cal H}=\pm \eta {\cal H}\, ,\quad P_\pm  \partial_y S   =   \partial_y S   P_\mp \, . 
\end{aligned}
\end{eqnarray}
  
  \section{PST symmetry in the ${\cal N}=(0,1)$ case}
  We assume constant background fields and for notation convenience define, 
  \begin{equation}
  \sigma = \frac{1}{\partial_{=} f} P_{+} \partial_{=} \IX \ , \quad \rho = \frac{1}{\partial_{\pp\,} f} P_{-} \partial_{\pp} \IX \ .
  \end{equation}
  In terms of these quantities we can recast the Lagrangian as,
  \begin{eqnarray}
  \begin{aligned}
  {\cal L} =- i\partial_{\pp\,} f D_- X\eta \left( \sigma + \rho \right) + i D_-f \partial_{\pp\,}\IX \eta \left( \sigma + \rho  \right) +  i\Psi_{\pp\, + } \eta \mu_{=} \ , 
  \end{aligned}
  \end{eqnarray} 
where we have defined $\Psi_{\pp\, +} \partial_{=} f   = 2 \Psi_{+}$ and in which the constraint, and its derivative are given by, 
  \begin{equation}
  \mu_=  = D_-f P_+ D_- \IX \  , \quad \nu_- = \frac{2 i}{\partial_{=} f} D_- \mu_=  = P_+ D_-X - \sigma D_-f \ . 
  \end{equation}
  The PST transformations in the case of constant backgrounds  reduce to,
  \begin{equation}
  \delta f = \varepsilon  \ , \quad \delta \IX = \varepsilon (\sigma + \rho) \ ,
  \end{equation}
  which exactly replicate those already seen in the bosonic ${\cal N}=(0,0)$ case.  
  Under these transformation one finds,
 \begin{equation}\label{eq:deltaconst}
 \delta \mu_= =  D_-\left( \varepsilon \nu_- \right) \ ,
 \end{equation}
  and the variation of the Lagrangian reads,
  \begin{equation}
 \delta {\cal L} = i \delta \Psi_{\pp\, +}\eta \mu_= + \Lambda_{\pp\,} \eta\nu_- - \Lambda_{\pp\,\pp\,}\eta \partial_= \nu_- \ , 
  \end{equation}
 in which we defined 
 \begin{equation}
\begin{aligned} 
\Lambda_\pp &=  i \varepsilon\left(D_-\Psi_{\pp\,+} - \frac{\partial_{\pp\,}f}{\partial_= f} \partial_= \sigma + \partial_{\pp\,}\sigma \right) \ ,\\
\Lambda_{\pp\,\pp\,} &=   - i \varepsilon \left( \frac{\partial_{\pp\,} f}{\partial_= f}\sigma - \frac{\partial_{\pp\,} \IX}{\partial_= f} \right) \ . 
 \end{aligned}
 \end{equation}
 Then   invariance of the action is recovered with, 
 \begin{equation}
\delta \Psi_{\pp\, +} = D_-\left( \frac{2 \Lambda_{\pp\,} }{\partial_= f} \right) + D_-\left( \frac{2  }{\partial_= f} \partial_=  \Lambda_{\pp\,\pp\,} \right) \ . 
 \end{equation}
One could choose other rewritings of the action by adding on terms proportional to the constraint, but due to eq.~\eqref{eq:deltaconst}   the transformation rule of the Lagrange multiplier can be modified to ensure the resulting action still possesses the PST symmetry. 
 
\end{appendix}

%
%


\begin{thebibliography}{99}

\bibitem{Duff:1989tf}
  M.~J.~Duff,
{\em Duality Rotations in String Theory},
  Nucl.\ Phys.\ B {\bf 335} (1990) 610.
  doi:10.1016/0550-3213(90)90520-N

\bibitem{Tseytlin:1990nb}
  A.~A.~Tseytlin,
{\em Duality Symmetric Formulation Of String worldsheet Dynamics},
  Phys.\ Lett.\ B {\bf 242} (1990) 163.  
\bibitem{Tseytlin:1990va}
  A.~A.~Tseytlin,
{\em Duality symmetric closed string theory and interacting chiral scalars},
  Nucl.\ Phys.\ B {\bf 350} (1991) 395.
\cite{Hull:2004in}

\bibitem{Hull:2004in}
  C.~M.~Hull,
{\em A Geometry for non-geometric string backgrounds},
  JHEP {\bf 0510} (2005) 065
  [hep-th/0406102].

\bibitem{Hull:2006va}
  C.~M.~Hull,
{\em Doubled Geometry and T-Folds},
  JHEP {\bf 0707} (2007) 080
  doi:10.1088/1126-6708/2007/07/080
  [hep-th/0605149].

\bibitem{Siegel:1993th}
  W.~Siegel,
{\em Superspace duality in low-energy superstrings},
  Phys.\ Rev.\ D {\bf 48} (1993) 2826
  [hep-th/9305073].

%
\bibitem{Siegel:1993xq}
  W.~Siegel,
{\em Two vierbein formalism for string inspired axionic gravity},
  Phys.\ Rev.\ D {\bf 47} (1993) 5453
  [hep-th/9302036].

\bibitem{Hull:2009mi}
  C.~Hull and B.~Zwiebach,
{\em Double Field Theory},
  JHEP {\bf 0909} (2009) 099
  [arXiv:0904.4664 [hep-th]].



\bibitem{Berman:2010is}
  D.~S.~Berman and M.~J.~Perry,
{\em Generalized Geometry and M theory},
  JHEP {\bf 1106} (2011) 074
  [arXiv:1008.1763 [hep-th]].
 
\bibitem{Hohm:2013pua}
  O.~Hohm and H.~Samtleben,
{\em Exceptional Form of D=11 Supergravity},
  Phys.\ Rev.\ Lett.\  {\bf 111} (2013) 231601
  doi:10.1103/PhysRevLett.111.231601
  [arXiv:1308.1673 [hep-th]].

\bibitem{West:2001as}
  P.~C.~West,
{\em E(11) and M theory},
  Class.\ Quant.\ Grav.\  {\bf 18} (2001) 4443
  [hep-th/0104081].
  
\bibitem{Aldazabal:2013sca}
  G.~Aldazabal, D.~Marques and C.~Nunez,
{\em Double Field Theory: A Pedagogical Review},
  Class.\ Quant.\ Grav.\  {\bf 30} (2013) 163001
  [arXiv:1305.1907 [hep-th]].
  
\bibitem{Hohm:2013bwa}
  O.~Hohm, D.~L\"ust and B.~Zwiebach,
{\em The Spacetime of Double Field Theory: Review, Remarks, and Outlook},
  Fortsch.\ Phys.\  {\bf 61} (2013) 926
  [arXiv:1309.2977 [hep-th]].

\bibitem{Berman:2013eva}
  D.~S.~Berman and D.~C.~Thompson,
{\em Duality Symmetric String and M-Theory},
  Phys.\ Rept.\  {\bf 566} (2014) 1
  [arXiv:1306.2643 [hep-th]].
  
  
  
\bibitem{Freidel:2013zga}
  L.~Freidel, R.~G.~Leigh and D.~Minic,
{\em Born Reciprocity in String Theory and the Nature of Spacetime},
  Phys.\ Lett.\ B {\bf 730} (2014) 302
  [arXiv:1307.7080].
  
\bibitem{Freidel:2015pka}
  L.~Freidel, R.~G.~Leigh and D.~Minic,
{\em Metastring Theory and Modular Space-time},
  arXiv:1502.08005 [hep-th].
  
  
\bibitem{Hohm:2011dv}
  O.~Hohm, S.~K.~Kwak and B.~Zwiebach,
{\em Double Field Theory of Type II Strings},
  JHEP {\bf 1109} (2011) 013
  [arXiv:1107.0008 [hep-th]].
  
  
\bibitem{Hitchin:2004ut}
  N.~Hitchin,
  {\em Generalized Calabi-Yau manifolds}, {\em
  Quart.\ J.\ Math.\ Oxford Ser.\ } {\bf 54}, 281 (2003)
  {\tt [arXiv:math/0209099]};
  
  
%
\bibitem{Gualtieri:2003dx}
  M.~Gualtieri,
{\em Generalized complex geometry}, Oxford University DPhil thesis, 2003,
{\tt arXiv:math/0401221};

%
 

\bibitem{Coimbra:2011nw}
  A.~Coimbra, C.~Strickland-Constable and D.~Waldram,
{\em Supergravity as Generalised Geometry I: Type II Theories},
  JHEP {\bf 1111} (2011) 091
  [arXiv:1107.1733 [hep-th]].

\bibitem{Coimbra:2012af}
  A.~Coimbra, C.~Strickland-Constable and D.~Waldram,
 {\em Supergravity as Generalised Geometry II: $E_{d(d)} \times \mathbb{R}^+$ and M theory},
  JHEP {\bf 1403} (2014) 019
  [arXiv:1212.1586 [hep-th], arXiv:1212.1586].
 
  
  


\bibitem{Hohm:2011cp}
  O.~Hohm and S.~K.~Kwak,
{\em Massive Type II in Double Field Theory},
  JHEP {\bf 1111} (2011) 086
  [arXiv:1108.4937 [hep-th]].
  
\bibitem{Geissbuhler:2011mx}
  D.~Geissbuhler,
 {\em Double Field Theory and N=4 Gauged Supergravity},
  JHEP {\bf 1111} (2011) 116
  doi:10.1007/JHEP11(2011)116
  [arXiv:1109.4280 [hep-th]].
\bibitem{Aldazabal:2011nj}
  G.~Aldazabal, W.~Baron, D.~Marques and C.~Nunez,
  {\em The effective action of Double Field Theory},
  JHEP {\bf 1111} (2011) 052
   Erratum: [JHEP {\bf 1111} (2011) 109]
  doi:10.1007/JHEP11(2011)052, 10.1007/JHEP11(2011)109
  [arXiv:1109.0290 [hep-th]].
\bibitem{Grana:2012rr}
  M.~Grana and D.~Marques,
 {\em Gauged Double Field Theory},
  JHEP {\bf 1204} (2012) 020
  doi:10.1007/JHEP04(2012)020
  [arXiv:1201.2924 [hep-th]].
\bibitem{Berman:2012uy}
  D.~S.~Berman, E.~T.~Musaev, D.~C.~Thompson and D.~C.~Thompson,
{\em Duality Invariant M-theory: Gauged supergravities and Scherk-Schwarz reductions},
  JHEP {\bf 1210} (2012) 174
  doi:10.1007/JHEP10(2012)174
  [arXiv:1208.0020 [hep-th]].
  
  




  
\bibitem{HackettJones:2006bp}
  E.~Hackett-Jones and G.~Moutsopoulos,
{\em Quantum mechanics of the doubled torus},
  JHEP {\bf 0610} (2006) 062
  [hep-th/0605114].
  
\bibitem{Berman:2007vi}
  D.~S.~Berman and N.~B.~Copland,
{\em The String partition function in Hull's doubled formalism},
  Phys.\ Lett.\ B {\bf 649} (2007) 325
  [hep-th/0701080].
  
\bibitem{Tan:2014mba}
  H.~S.~Tan,
{\em Closed String Partition Functions in Toroidal Compactifications of Doubled Geometries},
  JHEP {\bf 1405} (2014) 133
  [arXiv:1403.4683 [hep-th]].
  
  
\bibitem{Floreanini:1987as}
  R.~Floreanini and R.~Jackiw,
{\em Selfdual Fields as Charge Density Solitons},
  Phys.\ Rev.\ Lett.\  {\bf 59} (1987) 1873.
  
\bibitem{Berman:2007xn}
  D.~S.~Berman, N.~B.~Copland and D.~C.~Thompson,
{\em Background Field Equations for the Duality Symmetric String},
  Nucl.\ Phys.\ B {\bf 791} (2008) 175
  [arXiv:0708.2267 [hep-th]].
\bibitem{Berman:2007yf}
  D.~S.~Berman and D.~C.~Thompson,
{\em Duality Symmetric Strings, Dilatons and O(d,d) Effective Actions},
  Phys.\ Lett.\ B {\bf 662} (2008) 279
  [arXiv:0712.1121 [hep-th]].
  

\bibitem{Copland:2011wx}
  N.~B.~Copland,
{\em A Double Sigma Model for Double Field Theory},
  JHEP {\bf 1204} (2012) 044
  [arXiv:1111.1828 [hep-th]].

\bibitem{Lee:2013hma}
  K.~Lee and J.~H.~Park,
{\em Covariant action for a string in ``doubled yet gauged'' spacetime},
  Nucl.\ Phys.\ B {\bf 880} (2014) 134
  [arXiv:1307.8377 [hep-th]].

\bibitem{Betz:2014aia}
  A.~Betz, R.~Blumenhagen, D.~L\"ust and F.~Rennecke,
{\em A Note on the CFT Origin of the Strong Constraint of DFT},
  JHEP {\bf 1405} (2014) 044
  [arXiv:1402.1686 [hep-th]].
%
%

\bibitem{Pasti:1996vs}
  P.~Pasti, D.~P.~Sorokin and M.~Tonin,
 {\em On Lorentz invariant actions for chiral p forms},  Phys.\ Rev.\ D {\bf 55} (1997) 6292  [hep-th/9611100].  


\bibitem{Rocek:1997hi}
  M.~Rocek and A.~A.~Tseytlin,
 {\em Partial breaking of global D = 4 supersymmetry, constrained superfields, and three-brane actions},
  Phys.\ Rev.\ D {\bf 59} (1999) 106001
  [hep-th/9811232].

\bibitem{Nibbelink:2012jb}
  S.~Groot Nibbelink and P.~Patalong,
{\em A Lorentz invariant doubled world-sheet theory},
  Phys.\ Rev.\ D {\bf 87} (2013) 4,  041902
  [arXiv:1207.6110 [hep-th]].

\bibitem{Nibbelink:2013zda}
  S.~Groot Nibbelink, F.~Kurz and P.~Patalong,
{\em Renormalization of a Lorentz invariant doubled worldsheet theory},
  JHEP {\bf 1410} (2014) 114
  [arXiv:1308.4418 [hep-th]].

\bibitem{de la Ossa:1992vc}
  X.~C.~de la Ossa and F.~Quevedo,
{\em Duality symmetries from nonAbelian isometries in string theory},
  Nucl.\ Phys.\ B {\bf 403} (1993) 377
  [hep-th/9210021].

\bibitem{Klimcik:1995ux}
  C.~Klimcik and P.~Severa,
{\em Dual nonAbelian duality and the Drinfeld double},
  Phys.\ Lett.\ B {\bf 351} (1995) 455
  [hep-th/9502122].
\bibitem{Klimcik:1995dy}
  C.~Klimcik and P.~Severa,
{\em Poisson-Lie T duality and loop groups of Drinfeld doubles},
  Phys.\ Lett.\ B {\bf 372} (1996) 65
  doi:10.1016/0370-2693(96)00025-1
  [hep-th/9512040].
  

%
%

\bibitem{Chowdhury:2007ba}
  S.~P.~Chowdhury,
 {\em Superstring partition functions in the doubled formalism},
  JHEP {\bf 0709} (2007) 127
  [arXiv:0707.3549 [hep-th]].

\bibitem{Blair:2013noa}
  C.~D.~A.~Blair, E.~Malek and A.~J.~Routh,
  {\em An $O(D, D)$ invariant Hamiltonian action for the superstring},
  Class.\ Quant.\ Grav.\  {\bf 31} (2014) 20,  205011
  doi:10.1088/0264-9381/31/20/205011
  [arXiv:1308.4829 [hep-th]].

\bibitem{Rocek:1991ps}
  M.~Ro\v cek and E.~P.~Verlinde,
  {\em Duality, quotients, and currents},
  Nucl.\ Phys.\ B {\bf 373} (1992) 630
  doi:10.1016/0550-3213(92)90269-H
  [hep-th/9110053].

\bibitem{Buscher:1987sk}
  T.~H.~Buscher,
{\em A Symmetry of the String Background Field Equations},
  Phys.\ Lett.\ B {\bf 194} (1987) 59.
\bibitem{Buscher:1987qj}
  T.~H.~Buscher,
{\em Path Integral Derivation of Quantum Duality in Nonlinear Sigma Models},
  Phys.\ Lett.\ B {\bf 201} (1988) 466.


 \bibitem{Patalong:2013iya}
  P.~Patalong,
 {\em Aspects of non-geometry in string theory}, PhD thesis, Univ. Munich, Nov. 2013.

\bibitem{Sevrin:2013nca}
  A.~Sevrin and D.~C.~Thompson,
{\em A Note on Supersymmetric Chiral Bosons},
  JHEP {\bf 1307} (2013) 086
  [arXiv:1305.4848 [hep-th]].
  
  \bibitem{Giveon:1993ai}
  A.~Giveon and M.~Rocek,
{\em On nonAbelian duality},
  Nucl.\ Phys.\ B {\bf 421} (1994) 173
  doi:10.1016/0550-3213(94)90230-5
  [hep-th/9308154].
  
  
\bibitem{Alvarez:1994np}
  E.~Alvarez, L.~Alvarez-Gaume and Y.~Lozano,
  {\em On nonAbelian duality},
  Nucl.\ Phys.\ B {\bf 424} (1994) 155
  doi:10.1016/0550-3213(94)90093-0
  [hep-th/9403155].
\bibitem{Elitzur:1994ri}
  S.~Elitzur, A.~Giveon, E.~Rabinovici, A.~Schwimmer and G.~Veneziano,
{\em Remarks on nonAbelian duality},
  Nucl.\ Phys.\ B {\bf 435} (1995) 147
  doi:10.1016/0550-3213(94)00426-F
  [hep-th/9409011].

  
  
\bibitem{Cherkis:1997bx}
  S.~A.~Cherkis and J.~H.~Schwarz,
  {\em Wrapping the M theory five-brane on K3},
  Phys.\ Lett.\ B {\bf 403} (1997) 225
  [hep-th/9703062].
 
\bibitem{Giveon:1991jj}
  A.~Giveon and M.~Ro\v cek,
  ``Generalized duality in curved string backgrounds,''
  Nucl.\ Phys.\ B {\bf 380} (1992) 128
  doi:10.1016/0550-3213(92)90518-G
  [hep-th/9112070].
 
\bibitem{Lechner:1998ga}
  K.~Lechner,
 {\em Selfdual tensors and gravitational anomalies in 4n + 2-dimensions},  Nucl.\ Phys.\ B {\bf 537} (1999) 361  [hep-th/9808025].  


\bibitem{Sonnenschein:1988ug}
  J.~Sonnenschein,
{\em Chiral Bosons},
  Nucl.\ Phys.\ B {\bf 309} (1988) 752.

\bibitem{KSnotes}
K.~Sfetsos, private communication  of {\em Notes on Poisson-Lie T-duality} (1997)



\bibitem{Rocek:1978nb}
  M.~Ro\v cek,
{\em Linearizing the Volkov-Akulov Model},
  Phys.\ Rev.\ Lett.\  {\bf 41} (1978) 451.
  doi:10.1103/PhysRevLett.41.451;
U.~Lindstrom and M.~Ro\v cek,
{\em Constrained Local Superfields},
  Phys.\ Rev.\ D {\bf 19} (1979) 2300.
  doi:10.1103/PhysRevD.19.2300;
a systematic treatment and review of nilpotent superfields can be found in G.~Dall’Agata, E.~Dudas and F.~Farakos,
{\em On the origin of constrained superfields},
  JHEP {\bf 1605} (2016) 041
  doi:10.1007/JHEP05(2016)041
  [arXiv:1603.03416 [hep-th]].
  
\bibitem{AS}
A.~ Sevrin, {\em Some comments on supersymmetry and the doubled formalism from a worldsheet perspective}, Talk at 
Generalized Geometry and T-dualities, Simons Center For Geometry and Physics, May 2010.
\href{http://scgp.stonybrook.edu/video_portal/results.php?event_id=116}{{\tt{http://scgp.stonybrook.edu/video\_portal/results.php?event\_id=116}}}

\bibitem{Drinfeld:1986in}
  V.~G.~Drinfeld,
{\em Quantum groups},
  J.\ Sov.\ Math.\  {\bf 41} (1988) 898
   [Zap.\ Nauchn.\ Semin.\  {\bf 155} (1986) 18].
  doi:10.1007/BF01247086
  
  
\bibitem{Thompson:2015lzd}
  D.~C.~Thompson,
  {\em Generalised T-duality and Integrable Deformations},
  Fortsch.\ Phys.\  {\bf 64} (2016) 349
  doi:10.1002/prop.201500076
  [arXiv:1512.04732 [hep-th]].
  
\bibitem{Hull:2007jy}
  C.~M.~Hull and R.~A.~Reid-Edwards,
 {\em Gauge symmetry, T-duality and doubled geometry},
  JHEP {\bf 0808} (2008) 043
  doi:10.1088/1126-6708/2008/08/043
  [arXiv:0711.4818 [hep-th]].
  
\bibitem{Hull:2009sg}
  C.~M.~Hull and R.~A.~Reid-Edwards,
 {\em Non-geometric backgrounds, doubled geometry and generalised T-duality},
  JHEP {\bf 0909} (2009) 014
  doi:10.1088/1126-6708/2009/09/014
  [arXiv:0902.4032 [hep-th]].



  
  
\bibitem{Dall'Agata:2008qz}
  G.~Dall'Agata and N.~Prezas,
 {\em Worldsheet theories for non-geometric string backgrounds},
  JHEP {\bf 0808} (2008) 088
  doi:10.1088/1126-6708/2008/08/088
  [arXiv:0806.2003 [hep-th]].

\bibitem{Avramis:2009xi}
  S.~D.~Avramis, J.~P.~Derendinger and N.~Prezas,
  {\em Conformal chiral boson models on twisted doubled tori and non-geometric string vacua},
  Nucl.\ Phys.\ B {\bf 827} (2010) 281
  doi:10.1016/j.nuclphysb.2009.11.003
  [arXiv:0910.0431 [hep-th]].
  
\bibitem{Bandos:2015cha}
  I.~Bandos,
{\em Superstring in doubled superspace},
  Phys.\ Lett.\ B {\bf 751} (2015) 408
  doi:10.1016/j.physletb.2015.10.081
  [arXiv:1507.07779 [hep-th]].
  
\bibitem{Berman:1998va}
  D.~Berman,
{\em M5 on a torus and the three-brane},
  Nucl.\ Phys.\ B {\bf 533} (1998) 317
  doi:10.1016/S0550-3213(98)80009-6
  [hep-th/9804115].
  
\bibitem{Bandos:2014bva}
  I.~Bandos,
  JHEP {\bf 1408} (2014) 048
  doi:10.1007/JHEP08(2014)048
  [arXiv:1406.5185 [hep-th]].
 
 \end{thebibliography}
\end{document}